\documentclass[conf]{new-aiaa}
\usepackage[utf8]{inputenc}

\usepackage{graphicx}
\usepackage{amsmath}
\usepackage[version=4]{mhchem}
\usepackage{siunitx}
\usepackage{latexsym}
\usepackage{amsmath}
\usepackage{longtable,tabularx}
\usepackage{arydshln}
\usepackage{subcaption}
\usepackage{booktabs}
\usepackage{comment}
\usepackage[ruled,vlined]{algorithm2e}
\usepackage{algorithmic}
\usepackage{newtxtext,newtxmath}

 {    
      \newtheorem{assumption}{Assumption}
      
      \newtheorem{remark}{Remark}
      
      \newtheorem{proposition}{Proposition}
       
}

\usepackage[colorinlistoftodos]{todonotes}

\title{Nonlinear Incremental Control for Flexible Aircraft 
Trajectory Tracking and Load Alleviation}

\author{Xuerui Wang\footnote{Assistant Professor, Department of Aerospace Structures and Materials, Department of Control and Operations, Faculty of Aerospace Engineering, Kluyverweg 1, 2629HS Delft, the Netherlands, X.Wang-6@tudelft.nl, AIAA Member.}, Tigran Mkhoyan\footnote{Ph.D. Candidate, Department of Aerospace Structures and Materials, Faculty of Aerospace Engineering, Kluyverweg 1, 2629HS Delft, the Netherlands, T.Mkhoyan@tudelft.nl, AIAA Student Member.} and Roeland de Breuker\footnote{Associate Professor, Department of Aerospace Structures and Materials, Faculty of Aerospace Engineering, Kluyverweg 1, 2629HS Delft, the Netherlands, R.DeBreuker@tudelft.nl, AIAA Senior Member.}}
\affil{Delft University of Technology, Faculty of Aerospace Engineering, \\ Kluyverweg 1, 2629 HS Delft, The Netherlands}

\begin{document}

\maketitle

\begin{abstract}
This paper proposes a nonlinear control architecture for flexible aircraft simultaneous trajectory tracking and load alleviation. By exploiting the control redundancy, the gust and maneuver loads are alleviated without degrading the rigid-body command tracking performance. The proposed control architecture contains four cascaded control loops: position control, flight path control, attitude control, and optimal multi-objective wing control. Since the position kinematics are not influenced by model uncertainties, the nonlinear dynamic inversion control is applied. On the contrary, the flight path dynamics are perturbed by both model uncertainties and atmospheric disturbances; thus the incremental sliding mode control is adopted. Lyapunov-based analyses show that this method can simultaneously reduce the model dependency and the minimum possible gains of conventional sliding mode control methods. Moreover, the attitude dynamics are in the strict-feedback form; thus the incremental backstepping sliding mode control is applied. Furthermore, a novel load reference generator is designed to distinguish the necessary loads for performing maneuvers from the excessive loads. The load references are realized by the inner-loop optimal wing controller, while the excessive loads are naturalized by flaps without influencing the outer-loop tracking performance. The merits of the proposed control architecture are verified by trajectory tracking tasks and gust load alleviation tasks in spatial von K\'arm\'an turbulence fields. 
\end{abstract}

\section{Introduction}
\lettrine{T}{he} pursuit of higher efficiency has driven aircraft design evolving towards having higher aspect ratios and more lightweight structures. The resulting increase in structural flexibility makes the conventional frequency separation between rigid-body and structural degrees of freedom unreliable. During aggressive flight maneuvers or in the presence of gusts, structural integrity is also challenged. Given this background, there has been increasing interest in flexible aircraft control algorithms that can achieve multiple objectives, which include traditional pilot command-following flight control, gust load alleviation (GLA), maneuver load alleviation (MLA), aeroelastic mode suppression, drag minimization, etc.

In the literature, it is a common practice to design two independent controllers, one for rigid-body command tracking and another for fulfilling the remaining objectives (GLA, MLA, flutter suppression, etc.)~\cite{Dillsaver2013,Drew2019a,Virtual2019a}. As reported in~\cite{Dillsaver2013}, conflict can occur if the roll tracking command and the load alleviation command are given to the same wing trailing-edge control surface. In order to avoid this conflict, the aerodynamic control surfaces are usually classified into two sets. For example, in~\cite{Dillsaver2013}, two inboard elevators are used for pitch control, two outboard elevators are used for roll control, while all the ailerons are used for GLA. A model prediction controller is designed for MLA in~\cite{Virtual2019a}; the outboard ailerons are used to achieve MLA, while the inboard ailerons together with elevator and rudder are selected for rigid-body control. A similar selection is also adopted in~\cite{Drew2019a} for executing the commands provided by linear-quadratic regulators. However, the selection and isolation cannot make full use of the control input space, and can cause a large gap between two adjacent flaps. An integrated controller that considers both rigid-body and aeroelastic control objectives, while utilizing the control redundancy is desirable. 

Another challenge faced by multi-objective flexible aircraft flight control algorithms is the balance between load alleviation performance and rigid-body command tracking performance. For example, the simulation results in~\cite{Nguyen2017} show that the pitch rate command tracking is poor because the MLA function is designed to reduce the wing root bending moment~\cite{Nguyen2017}. Essentially, if a flexible aircraft has more than three independent control surfaces, and if not all of them are in the aircraft symmetrical plane, then there are infinite deflection combinations to realize the conventional three-axis attitude control moments. In other words, such an aircraft is over-actuated in rigid-body command tracking tasks. Therefore, it is physically realistic to simultaneously achieve desired rigid-body control moments together with other objectives (such as load alleviation), rather than making trade-offs among them.  

Recently, a control method that can alleviate loads without changing rigid-body tracking performance is presented in~\cite{Hansena,Duan2019}. By exploiting the null space between the input and rigid-body output, the load alleviation objective is decoupled from the rigid-body tracking objective. However, this method has two main limitations: first, the design of the null space filters strongly depends on the matrix fraction description of the linear state-space model; thus the decoupling effectiveness is questioned in the presence of model uncertainties and unmodeled nonlinearities; second, the proposed controller only ensures the load bounds are not violated but does not further minimize the load variations within the bounds. These two issues will be tackled in this paper. 

Incremental control refers to a class of nonlinear control methods that replace a part of model information by sensor measurements. The most well-known incremental control method is the incremental nonlinear dynamic inversion (INDI) control, which was initially proposed in~\cite{Sieberling2010}. In the past decade, the reduced model-dependency and strong robustness of INDI have stimulated its application to various aerospace systems (a CS-25 certified aircraft~\cite{Grondman2018}, quadrotors~\cite{Sun2020}, launch vehicles~\cite{Mooij2020a}, etc.). Recently, Ref.~\cite{Wang2019b} presented more generalized derivations, as well as Lyapunov-based stability and robustness analyses for the INDI control. Theoretical analyses also show INDI is more robust than the nonlinear dynamic inversion used in~\cite{Shearer2008a}. The INDI control law has been used to solve a free-flying flexible aircraft GLA problem in~\cite{Wang2019c}. The control objectives of rigid-body motion regulation, GLA, and structural vibration suppression are achieved simultaneously. Moreover, theoretical analyses and various time-domain simulations demonstrate the robustness of INDI control to atmospheric disturbance, aerodynamic model uncertainties, and even actuator faults. Apart from these merits, the controller designed in~\cite{Wang2019c} also has limitations. Because the number of controlled states is higher than the number of control inputs, the weight-least square method is used in the INDI loop for making trade-offs among different virtual control channels. As a consequence, not all the designed virtual controls can be fully executed. Although this control structure works well for the GLA problem, it would degrade aircraft tracking performance if is directly extended to MLA problems. In view of this issue, a different control architecture will be proposed in this paper. 

To further expand the applicability and enhance the robustness of INDI control, the incremental sliding mode control framework (INDI-SMC) was proposed in~\cite{Wang2018b}. This hybrid framework is derived for generic multi-input/multi-output nonlinear systems and is able to simultaneously reduce the minimum possible control/observer gains as well as the model dependency of conventional SMC methods. Furthermore, theoretical proofs and simulations demonstrate that a wide range of model uncertainties, sudden actuator faults, and structural damage can be passively resisted by INDI-SMC, without using additional fault detection and isolation modules~\cite{Wang2018b}. The hybridization idea was carried forward in~\cite{Wang2019}, where the incremental backstepping sliding mode control (IBSMC) is proposed for multi-input/multi-output nonlinear strict-feedback perturbed systems. Theoretical analyses and simulations show IBSMC has less model dependency but enhanced robustness than model-based backstepping sliding mode control in the literature. These advantages make INDI-SMC and IBSMC promising for solving flexible aircraft control problems. 

The main contribution of this paper is a nonlinear control architecture designed for flexible aircraft trajectory tracking and load alleviation purposes. The proposed control architecture has the following features: 1) it has enhanced robustness against model uncertainties, external disturbances and faults despite its reduced model dependency; 2) it contains two load reference generation algorithms to distinguish the loads that are necessary to perform maneuvers from the excessive loads; 3) it can neutralize the excessive loads without degrading the rigid-body tracking performance, no matter the excessive loads are induced by maneuvers or atmospheric disturbances.

Apart from the control design, the kinematics and dynamics of a free-flying flexible aircraft are also derived in this paper. A body-fixed reference frame is used to capture both the inertial and aerodynamic couplings between the rigid-body and structural degrees of freedom. Furthermore, a modular approach to conveniently make an existing clamped-wing aeroservoelastic model free-flying is presented in this paper. This modular approach contributes to bridging the flight dynamic and aeroelasticity communities.  

The rest of this paper is structured as follows. The flexible aircraft kinematics and dynamics are derived in Sec.~\ref{sec_model}. The nonlinear control architecture is presented in Sec.~\ref{sec_control}. The simulation results are presented in Sec.~\ref{sec_results} with the conclusions drawn in Sec.~\ref{sec_conclusions}. In this paper, bold symbols represent vectors and matrices.

\section{Model Descriptions}
\label{sec_model}
\subsection{Flight Kinematics}

\begin{figure}[!h]
  \centering
  \includegraphics[width=0.85\textwidth]{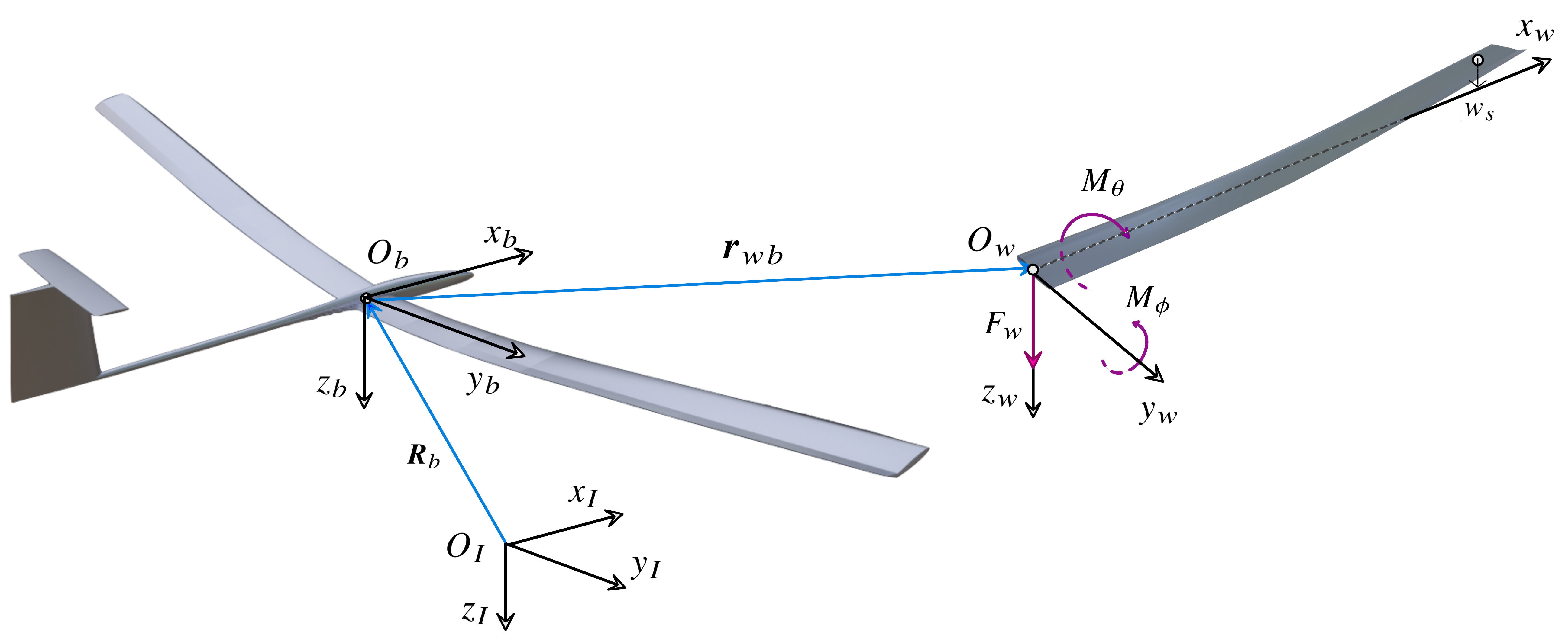}
  \caption{Reference frames and axis definitions of the flexible aircraft.}
  \label{fig:ref_frames}
\end{figure}

An overview of the flexible aircraft model used in this paper is shown in Fig.~\ref{fig:ref_frames}. 
To begin with, the following reference frames are defined:
\begin{itemize}
    \item Inertia Frame: $\mathcal{F}_I$ ($O_I,x_I,y_I,z_I$). Under the assumption that the Earth is flat and non-rotating, the Earth-center Earth-fixed reference frame can be used as an inertia frame.
    \item Body-fixed Reference Frame: $\mathcal{F}_B$ ($O_b,x_b,y_b,z_b$). In this paper, $O_b$ is selected at a fixed point on the fuselage, which is \textit{not necessarily} the center of mass (c.m.). This more general approach is adopted because during flight operations, c.m. is time-varying due to structural deformations and fuel assumptions. $O_bx_b$ and $O_bz_b$ are defined in the undeformed aircraft symmetrical plane.
    \item Wing-fixed Reference Frame: $\mathcal{F}_W$ ($O_w,x_w,y_w,z_w$). The left and right wings are modeled as two Euler Bernoulli beams. $O_w$ is defined at the root of each beam, and the distance vector from $O_b$ to $O_w$ is denoted as $\boldsymbol{r}_{wb}$. $O_wx_w$ is aligned with the \textit{undeformed} beam, and $O_wy_w$ is aligned with the mean-chord.
    \item Flight Trajectory Axes: $\mathcal{F}_V$ ($O_b,x_V,y_V,z_V$). $O_bx_V$ is aligned with the aircraft ground velocity vector, and $O_bz_V$ pointing downwards in the plumb plane.
    \item Aerodynamic Axes: $\mathcal{F}_A$ ($O_b,x_A,y_A,z_A$). $O_bx_A$ is aligned with the aircraft airspeed vector, and $O_bz_A$ pointing downwards in the aircraft symmetrical plane.
\end{itemize}

All the reference frames are right-handed. In Fig.~\ref{fig:ref_frames}, $F_w$ represents the wing root shear force. $M_\theta$ and $M_\phi$ respectively represents the wing root torsion and bending moment. 

Define the distance vector from $O_I$ to $O_b$ expressed in the inertial frame as $\boldsymbol{R}_b = [X,Y,Z]^\mathsf{T}$, then the translational kinematics are
\begin{equation}
    \begin{bmatrix}
   \dot X \\
   \dot Y\\
   \dot Z
    \end{bmatrix} = 
    \begin{bmatrix}
    V \cos \chi \cos \gamma \\
    V \sin \chi \cos \gamma \\
    -V \sin \gamma 
    \label{eq_position}
    \end{bmatrix}
\end{equation}
where $V$ is the ground velocity of the aircraft, $\gamma$ is the flight path angle, and $\chi$ is the kinematic azimuth angle.  

The attitude of an aircraft can be described by three Euler angles. Apart from them, for evaluating the influences of attitudes on aerodynamic forces, the following three attitude angles are more widely used: angle of attack $\alpha$, side-slip angle $\beta$, and the kinematic bank angle $\mu$. The attitude kinematics of an aircraft are given as follows~\cite{Stevens}:
\begin{equation}
    \begin{bmatrix}
    \dot \mu \\
    \dot \alpha \\
    \dot \beta
    \end{bmatrix} = 
    \begin{bmatrix}
    \cos \alpha \cos \beta & 0 & \sin \alpha \\
    \sin \beta & 1 & 0\\
    \sin \alpha\cos\beta & 0 &-\cos \alpha
    \end{bmatrix}^{-1}
    \left(
    -\boldsymbol{C}_{BV} 
    \begin{bmatrix}
    - \dot \chi \sin \gamma \\
    \dot \gamma \\
    \dot \chi \cos \gamma 
    \end{bmatrix} + 
    \begin{bmatrix}
    p\\q\\r
    \end{bmatrix}
    \right)
    \label{kine_atti}
\end{equation}

$\boldsymbol C_{ij}$ is used to denote the direction cosine matrix from the reference frame $\mathcal{F}_j$ to the reference frame $\mathcal{F}_i$. In this paper, only the left and right wings are modeled as flexible beams, while the remaining aircraft components are assumed to be rigid. 

\subsection{Aeroservoelastic Model of the Wing}
\label{subsection_ase_model}

In this subsection, the aeroservoelastic model of the wing will be derived in the local $\mathcal{F}_W$ frame. The right-wing dynamics will be derived as an example, while the dynamics of the left-wing can be derived analogously. The rigid-body motions influence the wing dynamics in two aspects: 1) change the local angle of attack of the wing; 2) cause additional inertial forces if rigid-body linear and angular accelerations are non-zero. Both the aerodynamic and inertial influences of rigid-body motions will be considered in the derivations. 

\subsubsection{Structural Model}

The wing structure is modeled as a linear dynamic Euler Bernoulli beam, which is discretized into $n_s$ elements with $n_s+1$ nodes. At each node, there are three degrees of freedom: transverse displacement $w_s$ (downwards positive), torsion $\theta_s$ (nose-up positive) and out-of plane bending $\phi_s$ (bend-up positive). A distributed trailing-edge control surface configuration is adopted in this paper. At each node, a flap with a deflection angle $\beta_s$ is connected to the beam through a rotational spring. $\beta_s >0$ denotes a flap downwards deflection. The inertia couplings between the flap and the beam are also considered. Define the state vector for this beam-flap system as $\boldsymbol{X}_s = [\boldsymbol{x}_{s,0}^\mathsf{T},...,\boldsymbol{x}_{s,n_s}^\mathsf{T}]^\mathsf{T}\in\mathbb{R}^{4(n_s+1)}$, where $\boldsymbol{x}_{s,i} = [w_{s,i},\phi_{s,i},\theta_{s,i},\beta_{s,i}]^\mathsf{T},~i=0,...,n_s$. Assume the beam is clamped at the first node, and denote $\boldsymbol{x}_s = [\boldsymbol{x}_{s,1}^\mathsf{T},...,\boldsymbol{x}_{s,n_s}^\mathsf{T}]^\mathsf{T} \in\mathbb{R}^{4n_s}$, then the corresponding dynamic equations are 
\begin{equation}
    \boldsymbol{M}_s \ddot {\boldsymbol{X}}_s + \boldsymbol{K}_s  \boldsymbol{X}_s =
    \left[
    \begin{array}{c:c}
     \boldsymbol{M}_{s,0} &  \boldsymbol{M}_{s,1} \\ \hdashline
     \boldsymbol{M}_{s,1}^\mathsf{T} & \boldsymbol{M}_{s,n_s}  
    \end{array}
     \right]
     \begin{bmatrix}
     \ddot {\boldsymbol{x}}_{s,0} \\ \hdashline
     \ddot {\boldsymbol{x}}_s
     \end{bmatrix}
     +
     \left[\begin{array}{c:c}
     \boldsymbol{K}_{s,0} &  \boldsymbol{K}_{s,1} \\ \hdashline
     \boldsymbol{K}_{s,1}^\mathsf{T} & \boldsymbol{K}_{s,n_s}  
    \end{array}
      \right]
     \begin{bmatrix}
      {\boldsymbol{x}}_{s,0} \\ \hdashline
      {\boldsymbol{x}}_s
     \end{bmatrix}
    =
    \left[
    \begin{array}{c}
      \boldsymbol{F}_{\text{root}}
        \\ \hdashline
     \boldsymbol{F}_{\text{ext}}
    \end{array}
    \right]
\label{eq_str}
\end{equation}

In Eq.~(\ref{eq_str}), $\boldsymbol{M}_s$ and $\boldsymbol{K}_s$ respectively represents the structure mass and stiffness matrices. The structural damping is neglected in this model. $\boldsymbol{F}_{\text{root}}$ is the reaction force at the clamped wing root. $\boldsymbol{F}_{\text{ext}}$ is the distributed external force vector, which contains gravitational and aerodynamic forces. Moreover, when the aircraft has linear and angular accelerations, the wing frame $\mathcal{F}_W$ becomes a non-inertial frame. Therefore, when modeling the structural dynamics in the wing frame, the inertial forces induced by rigid-body motions should be added to $\boldsymbol{F}_{\text{ext}}$. This inertial coupling effect is not modeled in the conventional mean-axes method~\cite{WASZAK1988}.  
\subsubsection{Inertial Forces}

Consider an infinitesimal mass element dm on an arbitrary section $A_w$ of the right wing. When the wing is undeformed, the distance vector from $O_{w}$ to dm, expressed in the right wing reference frame is $\boldsymbol{r}_w = [r_x,r_y,r_z]^\mathsf{T}$. Denote the transverse displacement of this wing section as $w$, and its torsion angle around the $O_{w}x_{w}$ axis as $\theta$, then the position vector caused by deformation is
\begin{eqnarray}
\boldsymbol{r}_e = 
\begin{bmatrix}
{r}_{e,x}\\{r}_{e,y}\\{r}_{e,z}
\end{bmatrix}  = 
\begin{bmatrix}
0\\0\\w
\end{bmatrix} + 
\begin{bmatrix}
0\\r_y  \cos\theta \\ r_y  \sin\theta 
\end{bmatrix}
\end{eqnarray}

The absolute distance from the inertial frame origin $O_I$ to $O_{w}$ equals the summation of $ \boldsymbol R_b$ (defined in the inertial frame) and $\boldsymbol{r}_{wb}$ (defined in the body reference frame). Denote $\boldsymbol{R}_{w}$ as the absolute distance from $O_I$ to dm, projected on the right wing reference frame, then
\begin{equation}
    \boldsymbol{R}_{w} =\boldsymbol{C}_{WB} \boldsymbol{C}_{BI} \boldsymbol R_b + \boldsymbol{C}_{WB}\boldsymbol{r}_{wb} +\boldsymbol{r}_w + \boldsymbol{r}_e 
\end{equation}

By differentiating the above equation, the absolute velocity of dm expressed in the right wing reference frame equals
\begin{eqnarray}
  \boldsymbol{V}_{w} &=& \boldsymbol{C}_{WB} \boldsymbol V_b +
     \boldsymbol{C}_{WB} \{\boldsymbol{\omega}_b\times[\boldsymbol{r}_{wb} +\boldsymbol{C}_{WB}^\mathsf{T}(  \boldsymbol{r}_w + \boldsymbol{r}_e ) ] \} + \boldsymbol{v}_e \nonumber \\
     &=& \boldsymbol{C}_{WB} \boldsymbol V_b +
     \boldsymbol{C}_{WB} \Tilde{\boldsymbol{\omega}}_b \boldsymbol{r}_{wb}+  \boldsymbol{C}_{WB} \Tilde{\boldsymbol{\omega}}_b\boldsymbol{C}_{WB}^\mathsf{T}(  \boldsymbol{r}_w + \boldsymbol{r}_e )  + \boldsymbol{v}_e
     \label{eq_velocity}
\end{eqnarray}
in which $\boldsymbol{V}_b = [V_x,V_y,V_z]^\mathsf{T}$ and $\boldsymbol{\omega}_b = [p,q,r]^\mathsf{T}$ are the translational and rotational velocities of the body-frame, and both of them are expressed in $\mathcal{F}_B$. $\boldsymbol{v}_e$ is the relative deformation velocity of dm, which equals $[\dot{r}_{e,x},\dot{r}_{e,y},\dot{r}_{e,z}]^\mathsf{T}$. $(\Tilde{\cdot})$ denotes the skew-symmetric matrix of the vector $({\cdot})$. 

Furthermore, differentiate Eq.~\eqref{eq_velocity}, the absolute acceleration of dm, expressed in the right wing reference frame equals
\begin{eqnarray}
  \boldsymbol{a}_{w} &=& \boldsymbol{C}_{WB} \boldsymbol a_f +
     \boldsymbol{C}_{WB}\Tilde{\boldsymbol{\alpha}}_f \boldsymbol{r}_{wb} + \boldsymbol{C}_{WB}\Tilde{\boldsymbol{\alpha}}_f \boldsymbol{C}_{WB}^\mathsf{T}(  \boldsymbol{r}_w + \boldsymbol{r}_e ) + 2 \boldsymbol{C}_{WB}\Tilde{\boldsymbol{\omega}}_b \boldsymbol{C}_{WB}^\mathsf{T} \boldsymbol{v}_e \nonumber \\
     &&+ \boldsymbol{a}_e + \boldsymbol{C}_{WB}\Tilde{\boldsymbol{\omega}}_b \Tilde{\boldsymbol{\omega}}_b  \boldsymbol{r}_{wb} + \boldsymbol{C}_{WB}\Tilde{\boldsymbol{\omega}}_b \Tilde{\boldsymbol{\omega}}_b \boldsymbol{C}_{WB}^\mathsf{T}(\boldsymbol{r}_w + \boldsymbol{r}_e)
     \label{eq_acc}
\end{eqnarray}
where $\boldsymbol a_b = [a_x,a_y,a_z]^\mathsf{T}$ and $\boldsymbol{\alpha}_b = [\dot p,\dot q,\dot r]^\mathsf{T}$ are the translational and rotational accelerations of the body-frame, and both of them are expressed in $\mathcal{F}_B$. $ \boldsymbol{a}_e = [\ddot{r}_{e,x},\ddot{r}_{e,y},\ddot{r}_{e,z}]^\mathsf{T} $ is the relative deformation acceleration of dm. From Eq.~\eqref{eq_acc}, the acceleration component of dm that is purely induced by rigid-body motion equals
\begin{equation}
     \boldsymbol{a}_{w}^{R} = \boldsymbol{C}_{WB} \boldsymbol a_f +
     \boldsymbol{C}_{WB}\Tilde{\boldsymbol{\alpha}}_f (\boldsymbol{r}_{wb} + \boldsymbol{C}_{WB}^\mathsf{T}\boldsymbol{r}_w )  + \boldsymbol{C}_{WB}\Tilde{\boldsymbol{\omega}}_b \Tilde{\boldsymbol{\omega}}_b  (\boldsymbol{r}_{wb}+ \boldsymbol{C}_{WB}^\mathsf{T}\boldsymbol{r}_w)
     \label{eq_acc_rigid}
\end{equation}

Integrate Eq.~\eqref{eq_acc_rigid} at the wing section $A_w$, then the inertial force purely induced by rigid-body motion per unit length is computed as 
\begin{equation}
    \boldsymbol{f}_{\text{acc}} = -\iint_{A_w} \rho \boldsymbol{a}_{w}^{R}~ \text{d} r_y \text{d} r_z
\end{equation}
where $\rho$ is the volume density. The inertial moment around $O_wx_w$ that is purely induced by rigid-body motion is 
\begin{equation}
    \boldsymbol{m}_{\text{acc}} = -\iint_{A_w} \rho [0,r_y,r_z]^\mathsf{T}\times \boldsymbol{a}_{w}^{R} ~\text{d} r_y  \text{d} r_z
\end{equation}

The distributed inertial forces and moments are integrated to their nearest structural node. Consider the $i$-th structural node with a spanwise location $r_x = x_i$, then the integrated force vector associated with its degrees of freedom is 
\begin{eqnarray}
\left[
\begin{array}{c}
     f_w  \\
     f_\phi \\
     f_\theta \\
     f_\beta
\end{array}
\right]_{\text{acc},i} = 
\begin{bmatrix}
\int_{x_{i-1}}^{x_i} {f}_{\text{acc},z} ~\text{d} r_x \\
\int_{x_{i-1}}^{x_i} (x_i - x) {f}_{\text{acc},z} -{m}_{\text{acc},x}  ~\text{d} r_x \\
\int_{x_{i-1}}^{x_i} {m}_{\text{acc},y} ~\text{d} r_x \\
0
\end{bmatrix}
\end{eqnarray}
which is valid for $i = 1,..,n_s$. The inertial loads on the structural nodes are collected into a force vector $\boldsymbol{F}_{\text{s,acc}}= [[ f_w,f_\phi, f_\theta,f_\beta]_{\text{acc},1},...,[ f_w,f_\phi, f_\theta,f_\beta]_{\text{acc},n_s}]^\mathsf{T} \in\mathbb R^{4n_s}$.  

\subsubsection{Aerodynamic Forces}

The unsteady strip theory is used to calculate the aerodynamic force $\boldsymbol{F}_{\text{aero}}$, which is caused by motions, flap deflections and external atmospheric disturbances. Discretize the wing into $n_a$ undeformable strips, where each one of them has three degrees of freedom: heave $w_a$ (downwards positive), pitching around the elastic axis $\theta_a$ (nose-up positive) and flap deflection $\beta_a$ (downwards positive). Referring to Theodorsen's theory, $\boldsymbol{F}_{\text{aero}}$ contains instant noncirculatory and time-dependent circulatory parts~\cite{Theodorsen1935}. The influences of noncirculatory force can be directly considered in the aerodynamic mass, damping and stiffness matrices. On the other hand, the circulatory part was initially expressed in frequency domain~\cite{Theodorsen1935}. For control design purposes, the circulatory aerodynamics are converted into time-domain using indicial function approximation.

Define a nondimensional time variable as $\tau = Vt/b$, where $b$ is the semi-chord. Using the Duhamel's integral, a dynamic system with indicial response function $f(\tau) = 1-a_1 e^{-b_1 \tau} -a_2 e^{-b_2 \tau}$ can be realized in a state-space form as
\begin{eqnarray}
\left[
\begin{array}{c}
\dot {z}_{a_1} \\
\dot {z}_{a_2} \\
\end{array}
\right]
&=&
\left[
\begin{array}{cc}
0 & 1 \\
-\left( \frac{V}{b} \right)^2 b_1 b_2  & -  \left( \frac{V}{b} \right)(b_1 +b_2) \\
\end{array}
\right]
\left[
\begin{array}{c}
{z}_{a_1} \\
{z}_{a_2} \\
\end{array}
\right] + 
\left[
\begin{array}{c}
0 \\
1 \\
\end{array}\right] u \nonumber \\
y &=& \left[ (a_1+a_2) b_1b_2\left( \frac{V}{b} \right)^2,~ (a_1b_1+a_2b_2) \left( \frac{V}{b} \right) \right]
\left[
\begin{array}{c}
{z}_{a_1} \\
{z}_{a_2} \\
\end{array}
\right]
+ (1-a_1-a_2) u
\label{eq_state_space}
\end{eqnarray}

The circulatory lift induced by aircraft motions and flap deflections is modeled by the Wagner's function, which has an exponential approximation $
\phi(\tau) = 1- 0.165 e^{-0.0455\tau} - 0.335 e^{-0.3\tau}$~\cite{aeroelasticity_book}. Since this function is in the form of $f(\tau)$, the state-space representation in Eq.~(\ref{eq_state_space}) can be applied. The input of the system is chosen as 
\begin{equation}
    u = \alpha_{{qs,r}}(\boldsymbol{V}_f,\boldsymbol{\omega}_b) + \alpha_{{qs,ae}} + \beta_{qs}
    \label{input_alpha}
\end{equation}
where the subscript $(\cdot)_{qs}$ indicates quasi-steady. From Eq.~\eqref{eq_velocity}, the local in-flow velocity that is purely caused by rigid-body motion equals $\boldsymbol{V}_w^R =  \boldsymbol{C}_{WB} \boldsymbol V_b +\boldsymbol{C}_{WB} \Tilde{\boldsymbol{\omega}}_b (\boldsymbol{r}_{wb}+  \boldsymbol{C}_{WB}^\mathsf{T}  \boldsymbol{r}_w ) $. Therefore, the local quasi-steady angle of attack induced by rigid-body translational and rotational motion equals $\alpha_{{qs,r}}(\boldsymbol{V}_b,\boldsymbol{\omega}_b) = \arctan({V}_{w,z}^R/{V}_{w,x}^R)$.

The angle of attack induced by aeroelastic motions is $\alpha_{{qs,ae}} = \frac{\dot{w}_a}{V} + \theta_a + b(\frac{1}{2}-a)\frac{\dot \theta_a}{V}$. Moreover, in Eq.~(\ref{input_alpha}), $\beta_{qs} =\frac{T_{10}}{\pi} \beta_a + \frac{T_{11}}{2\pi} \frac{b}{V}\dot{ \beta}_a $, where $T_{10}$ and $T_{11}$ can be found in~\cite{Theodorsen1935}. Substituting Eq.~(\ref{input_alpha}) and the parameters of the Wagner's function into Eq.~(\ref{eq_state_space}), then the output $y$ gives the effective time-dependent angle of attack. As a result, the local circulatory lift coefficient equals $ C_{L_\alpha}^{SF}y$. The Theodorsen's equations are derived for thin airfoils with lift-slope equals $2\pi$. In this paper, more general expressions are used, where the local $C_{L_\alpha}^{SF}$ is obtained from steady-flow analysis based on the vortex lattice method. This way allows the considerations of the wing-tip wake roll up effects and compressibility (Prandtl-Glauert factor) in the local $C_{L_\alpha}^{SF}$. Consequently, the resulting circulatory moment coefficient around the shear center equals $ C_{L_\alpha}^{SF}y(0.25+0.5a)$, and the circulatory hinge moment coefficient equals $-C_{L_\alpha}^{SF}y\frac{T_{12}}{4\pi}$ (the coefficient $T_{12}$ is given in~\cite{Theodorsen1935}).

The K\"ussner function is used to calculate the unsteady responses of an airfoil to an unit sharp-edged gust. One of its exponentially response approximation is $\psi_g(\tau) = 1- 0.5 e^{-0.13\tau} - 0.5e^{-\tau}$~\cite{aeroelasticity_book}. This expression is also in the form of $f(\tau) = 1-a_1 e^{-b_1 \tau} -a_2 e^{-b_2 \tau}$, thus the dynamic system in Eq.~(\ref{eq_state_space}) can be used to calculate the responses of an airfoil to an arbitrary-spectrum atmospheric disturbance. Consider a wing section encounters a vertical gust with velocity $w_g(t)$, then the gust velocity projected in the wing reference frame equals $ \boldsymbol V_{\text{gust}} =\boldsymbol{C}_{WB} \boldsymbol{C}_{BI}[0,0,w_g(t)]^\mathsf{T}$. Therefore, the input of Eq.~(\ref{eq_state_space}) is taken as $u = \alpha_g = \arctan(- V_{\text{gust},z}/V)$. Substituting the parameters of the K\"ussner function into Eq.~(\ref{eq_state_space}), then the output $y$ gives the effective time-dependent angle of attack induced by $w_g(t)$. The resulting lift, pitching moment and hinge moment coefficients are given by $C_{L_\alpha}^{SF} y$ multiplied with $1$, $0.25+0.5a$ and $-{T_{12}}/{4\pi}$, respectively.

Based on the above discussions, four additional states are needed by each strip to model the circulatory dynamics. These additional states are also known as lag states. Define the lag state vector as $\boldsymbol{z}_a\in \mathbb R^{4n_a}$, and define the aerodynamic state vector as $\boldsymbol{x}_a\in \mathbb R^{3n_a}$, then the following state-space system can be derived
\begin{equation}
    \dot {\boldsymbol{z}}_a  =\boldsymbol{A}_z \boldsymbol{z}_a + \boldsymbol{B}_z \boldsymbol{x}_a + \boldsymbol{B}_{\dot z} \dot{\boldsymbol{x}}_a + \boldsymbol{B}_{zr} \boldsymbol{\alpha}_{qs,r} + \boldsymbol{B}_{zg} \boldsymbol{\alpha}_{g}
    \label{eq_z}
\end{equation}
where $\boldsymbol{\alpha}_{qs,r} \in \mathbb R^{n_a}$ and $\boldsymbol{\alpha}_{g}\in \mathbb R^{n_a}$ are the angle of attack vectors induced by rigid-body motions and gusts. Define the aerodynamic output vector as $\boldsymbol{f}_{\text{aero}} = [\boldsymbol{f}_{\text{aero},1}^\mathsf{T},...,\boldsymbol{f}_{\text{aero},n_a}^\mathsf{T}]^\mathsf{T}$, where each element $\boldsymbol{f}_{\text{aero},i}$ contains lift, pitching moment around the elastic axis, and the hinge moment, then 
\begin{equation}
    \boldsymbol{f}_{\text{aero}} = \boldsymbol{M}_a \ddot {\boldsymbol{x}}_a + \boldsymbol{C}_a \dot {\boldsymbol{x}}_a + \boldsymbol{K}_a  {\boldsymbol{x}}_a + \boldsymbol{K}_z  {\boldsymbol{z}} + \boldsymbol{K}_{zr} \boldsymbol{\alpha}_{qs,r}
    \label{eq_aero}
\end{equation}

In the above equation, $\boldsymbol{M}_a$, $\boldsymbol{C}_a$, and $\boldsymbol{K}_a$ are the aerodynamic mass, damping and stiffness matrices, and they include both the noncirculatory and circulatory effects. Because in the Wagner's function, $1- a_1- a_2 = 0.5$, so the input of $\boldsymbol{\alpha}_{qs,r}$ has direct influence on $\boldsymbol{f}_{\text{aero}}$. On contrast, $1- a_1- a_2 = 0$ in the K\"ussner function, thus $\boldsymbol{\alpha}_{g}$ only indirectly influence $\boldsymbol{f}_{\text{aero}}$ though the lag state $\boldsymbol{z}$ (Eq.~\eqref{eq_z}). 


To establish a coupled aeroelastic model, the degrees of freedom of the aerodynamic strips and the structure nodes need to be connected. Normally, the number of strips $n_a$ is larger than the number of structure nodes $n_s$. Two linear interpolation matrices $\boldsymbol{H}_{as} \in \mathbb R^{3n_a\times 4n_s}$ and $\boldsymbol{H}_{sa} \in \mathbb R^{ 4n_s\times 3n_a}$ are designed based on the nearest neighbor principle. The aerodynamic degrees of freedom are interpolated from the structural states as $\boldsymbol{x}_a = \boldsymbol{H}_{as}  \boldsymbol{x}_s$. On the other hand, the distributed aerodynamic forces are assigned to the structural nodes as $\boldsymbol{F}_{s,\text{aero}} =\boldsymbol{H}_{sa} \boldsymbol{f}_{\text{aero}}$.


\subsubsection{Gravitational Forces}
   
The gravitational force per unit length expressed in the right-wing reference frame is 
\begin{equation}
    \boldsymbol{f}_{\text{grav}} = \iint_{A_w}  \boldsymbol{C}_{WB} \boldsymbol{C}_{BI} [0,0,\rho g]^\mathsf{T} ~ \text{d} r_x \text{d} r_z
\end{equation}   
where $g$ is the gravitational acceleration. The gravitational moment around $O_wx_w$ per unit length is 
\begin{equation}
    \boldsymbol{m}_{\text{grav}} = \iint_{A_w} ( [0,r_y,r_z]^\mathsf{T} +\boldsymbol{r}_e )\times \boldsymbol{C}_{WB} \boldsymbol{C}_{BI} [0,0,\rho g]^\mathsf{T} ~\text{d} r_x \text{d} r_z
\end{equation}

The distributed gravitational forces and moments are integrated to the nearest structural node. Consider the $i$-th structural node with spanwise location $r_x = x_i$, then the integrated force vector associated with its degrees of freedom is
\begin{eqnarray}
\left[
\begin{array}{c}
     f_w  \\
     f_\phi \\
     f_\theta \\
     f_\beta
\end{array}
\right]_{\text{grav},i} = 
\begin{bmatrix}
\int_{x_{i-1}}^{x_i} {f}_{\text{grav},z} ~\text{d} r_x \\
\int_{x_{i-1}}^{x_i} (x_i - x) {f}_{\text{grav},z} -{m}_{\text{grav},x}  ~\text{d} r_x \\
\int_{x_{i-1}}^{x_i} {m}_{\text{grav},y} ~\text{d} r_x \\
0
\end{bmatrix}
\end{eqnarray}
which is valid for $i = 1,..,n_s$. The gravitational loads at structural nodes are collected into a force vector $\boldsymbol{F}_{\text{s,grav}}= [[ f_w,f_\phi, f_\theta,f_\beta]_{\text{grav},1},...,[ f_w,f_\phi, f_\theta,f_\beta]_{\text{grav},n_s}]^\mathsf{T} \in\mathbb R^{4n_s}$.  

\subsubsection{Actuation Forces}

In this paper, flaps are attached to structural nodes via rotational springs. In Eq.~\eqref{eq_str}, the flap stiffness is considered in $\boldsymbol{K}_s$, while the flap mass and its inertial couplings with the beam structure are modeled in $\boldsymbol{M}_s$. The actuation moment around the hinge is chosen as the control input, which is more physically meaningful than the flap angle itself. Denote the actuation moment for the $i$-th flap as $u_i$, then the total actuation force vector is written as $\boldsymbol{F}_{s,\text{act}}  = [[0,0,0,u_1],...,[0,0,0,u_{n_s}]]^\mathsf{T}$. Denote the control input vector as $\boldsymbol{u} = [u_1,...,u_{n_s}]^\mathsf{T}\in \mathbb R^{n_s}$, then $\boldsymbol{F}_{s,\text{act}} $ also equals $\boldsymbol{H}_u \boldsymbol{u}$, where $\boldsymbol{H}_u\in \mathbb {R}^{4n_s \times n_s}$ is a boolean selection matrix.   

\subsubsection{Wing Aeroservoelastic Dynamic Equations}

After the derivations of inertial, aerodynamic, gravitational, and actuation forces, the wing aeroservoelastic model is ready to be assembled. Because the origin of the wing-fixed reference frame $\mathcal{F}_W$ is coincide with the first structure node, when observing the wing motions in $\mathcal{F}_W$, we have $\ddot {\boldsymbol{x}}_{s,0} = \dot {\boldsymbol{x}}_{s,0} = {\boldsymbol{x}}_{s,0} = 0$. The $\boldsymbol{F}_{\text{ext}}$ in Eq.~\eqref{eq_str} equals the summation of $\boldsymbol{F}_{s,\text{aero}}$, $ \boldsymbol{F}_{s,\text{acc}}$, $ \boldsymbol{F}_{s,\text{grav}}$, and $\boldsymbol{F}_{s,\text{act}}$. Substituting Eqs.~\eqref{eq_z} and~\eqref{eq_aero} into Eq.~\eqref{eq_str}, the following state-space system can be derived:
\begin{eqnarray}
  \label{eq_wing_aero}
  \begin{bmatrix}
  \ddot {\boldsymbol{x}}_{s} \\
  \dot {\boldsymbol{x}}_{s} \\
  \dot {\boldsymbol{z}}_a
  \end{bmatrix} &=& 
  \begin{bmatrix}
  \boldsymbol{M}_{ae}^{-1} \boldsymbol{H}_{sa} \boldsymbol{C}_a \boldsymbol{H}_{as}  &  \boldsymbol{M}_{ae}^{-1} (\boldsymbol{H}_{sa} \boldsymbol{K}_a \boldsymbol{H}_{as} - \boldsymbol{K}_{s,n_s}) & \boldsymbol{M}_{ae}^{-1} \boldsymbol{H}_{sa} \boldsymbol{K}_z  \\
  \boldsymbol{I}_{n_s\times n_s} & \boldsymbol{0}_{n_s\times n_s} & \boldsymbol{0}_{n_s\times 4n_a} \\
  \boldsymbol{B}_{\dot z}\boldsymbol{H}_{as} & \boldsymbol{B}_{ z}\boldsymbol{H}_{as}& \boldsymbol{A}_z
  \end{bmatrix}
  \begin{bmatrix}
  \dot {\boldsymbol{x}}_{s} \\
   {\boldsymbol{x}}_{s} \\
   {\boldsymbol{z}}_a
   \end{bmatrix}
  + 
  \begin{bmatrix}
  \boldsymbol{M}_{ae}^{-1}  \boldsymbol{H}_{u}\\
  \boldsymbol{0}_{n_s\times 1} \\
  \boldsymbol{0}_{4n_a\times 1}
  \end{bmatrix} \boldsymbol{u}  \\ \nonumber
  &&+
  \begin{bmatrix}
  \boldsymbol{0}_{n_s \times n_a} \\
  \boldsymbol{0}_{n_s \times n_a} \\
  \boldsymbol{B}_{zg}
  \end{bmatrix}  \boldsymbol{\alpha}_{g}+
  \begin{bmatrix}
  \boldsymbol{M}_{ae}^{-1}  \boldsymbol{F}_{s,\text{grav}}\\
  \boldsymbol{0}_{n_s\times 1} \\
  \boldsymbol{0}_{4n_a\times 1}
  \end{bmatrix}+
  \begin{bmatrix}
  \boldsymbol{M}_{ae}^{-1}\boldsymbol{H}_{sa} \boldsymbol{K}_{zr} \\
  \boldsymbol{0}_{n_s\times n_a} \\
  \boldsymbol{B}_{zr} 
  \end{bmatrix}  \boldsymbol{\alpha}_{qs,r} +
  \begin{bmatrix}
  \boldsymbol{M}_{ae}^{-1}  \boldsymbol{F}_{s,\text{acc}}\\
  \boldsymbol{0}_{n_s\times 1} \\
  \boldsymbol{0}_{4n_a\times 1}
  \end{bmatrix}
\end{eqnarray}
in which $\boldsymbol{M}_{ae} =\boldsymbol{M}_{s,n_s} - \boldsymbol{H}_{sa} \boldsymbol{M}_a \boldsymbol{H}_{as}$. Recall Eq.~\eqref{eq_str}, the wing root reaction force is calculated as
\begin{eqnarray}
      \boldsymbol{F}_{\text{root}} &=& \boldsymbol{M}_{s,1}\ddot {\boldsymbol{x}}_{s} + \boldsymbol{K}_{s,1} {\boldsymbol{x}}_{s} = \boldsymbol{M}_{s,1}\boldsymbol{M}_{ae}^{-1}\boldsymbol{H}_{sa} \boldsymbol{K}_{zr}\boldsymbol{\alpha}_{qs,r} + \boldsymbol{M}_{s,1}\boldsymbol{M}_{ae}^{-1} (\boldsymbol{B}_u \boldsymbol{u}+  \boldsymbol{F}_{s,\text{acc}} + \boldsymbol{F}_{s,\text{grav}}) \nonumber  \\
      && + \begin{bmatrix}
      \boldsymbol{M}_{s,1} \boldsymbol{M}_{ae}^{-1} \boldsymbol{H}_{sa} \boldsymbol{C}_a \boldsymbol{H}_{as} & \boldsymbol{M}_{s,1} \boldsymbol{M}_{ae}^{-1} (\boldsymbol{H}_{sa} \boldsymbol{K}_a \boldsymbol{H}_{as} - \boldsymbol{K}_{s,n_s}) + \boldsymbol{K}_{s,1} & \boldsymbol{M}_{s,1}\boldsymbol{M}_{ae}^{-1} \boldsymbol{H}_{sa} \boldsymbol{K}_z 
      \end{bmatrix} 
      \begin{bmatrix}
      \dot {\boldsymbol{x}}_{s} \\
   {\boldsymbol{x}}_{s} \\
   {\boldsymbol{z}}_a
      \end{bmatrix} 
      \label{eq_F_root}
\end{eqnarray}

The above wing aeroservoelastic model is derived in a modular approach. The control command, atmospheric disturbance, gravitational force, the angle of attack induced by rigid-body motions, and the inertial forces induced by rigid-body accelerations are modeled as separate inputs to the system. This allows convenient contribution assessments of different factors on the wing dynamic loads. Furthermore, this approach provides a clear interface between the conventional clamped-wing aeroservoelastic dynamics and the free-flying wing aeroservoelastic dynamics. If $\boldsymbol{F}_{s,\text{acc}}$ and $\boldsymbol{\alpha}_{qs,r}$ are set to zero, while the orientation matrices in $\boldsymbol{F}_{s,\text{grav}}$ are evaluated at the trimmed condition, then Eqs.~\eqref{eq_wing_aero} and \eqref{eq_F_root} degenerate to the conventional clamped-wing aeroservoelastic dynamic equations. On the other hand, no matter how the aerodynamic and structural dynamics are modeled and coupled (finite-element method, panel method, etc.), as long as the resulting aeroelastic model can be written in state-space form, the clamped-wing model can be made free-flying by adding $\boldsymbol{F}_{s,\text{acc}}$, $\boldsymbol{\alpha}_{qs,r}$ and the direction cosine matrices. 

\subsection{Free-flying Dynamics of the Flexible Aircraft}

The translational dynamics of the flexible aircraft expressed in the flight trajectory axes $\mathcal{F}_V$ are
\begin{equation}
    \begin{bmatrix}
    \dot V \\
    \dot \chi \\
    \dot \gamma
    \end{bmatrix} = 
    \begin{bmatrix}
    1 & 0&0\\
    0& \frac{1}{V \cos\gamma} & 0 \\
    0&0& -\frac{1}{V}
    \end{bmatrix} 
    \frac{\boldsymbol{C}_{VB}}{m} (\boldsymbol{F}_{\text{tot}}^B - \Tilde{\boldsymbol{S}}^\mathsf{T} \boldsymbol{\alpha}_b  - \Tilde{\boldsymbol{\omega}}_b \Tilde{\boldsymbol{S}}^\mathsf{T} {\boldsymbol{\omega}}_b + \boldsymbol{d}_F)
    \label{eq_Vxigamma}
\end{equation}
while the rotational dynamics of the aircraft expressed in the body-fixed frame $\mathcal{F}_B$ are 
\begin{equation}
    \boldsymbol{J} \dot{ \boldsymbol{\omega}}_b = - \Tilde{\boldsymbol{S}} \dot {\boldsymbol{V}}_b   - \Tilde{\boldsymbol{V}}_b \Tilde{\boldsymbol{S}}^\mathsf{T} \boldsymbol{\omega}_b
  - \Tilde{\boldsymbol{\omega}}_b \Tilde{\boldsymbol{S}}  {\boldsymbol{V}}_b - \Tilde{\boldsymbol{\omega}}_b \boldsymbol{J} {\boldsymbol{\omega}}_b + \boldsymbol{M}_{\text{tot}}^B +  \boldsymbol{d}_M
  \label{eq_dot_pqr}
\end{equation}

Denote the distance vector from $O_b$ to an infinitesimal mass element dm on the aircraft as $\boldsymbol{r}$, which is expressed in $\mathcal{F}_B$. Then ${\boldsymbol{S}} = \int {\boldsymbol{r}} ~\text {d} m$ is the first moment of area, which is non-zero when $O_b$ is not coincide with the center of mass. The moment of inertia is calculated as ${\boldsymbol{J}} = \int \Tilde{\boldsymbol{r}}^\mathsf{T}\Tilde{\boldsymbol{r}} ~\text {d} m$. For a flexible aircraft, both ${\boldsymbol{S}}$ and ${\boldsymbol{J}}$ are functions of structural deformations. 

$\boldsymbol{F}_{\text{tot}}^B$ and $\boldsymbol{M}_{\text{tot}}^B$ are the total external forces and moments expressed in $\mathcal{F}_B$, which contain aerodynamic, gravitational and propulsion forces. $\boldsymbol{d}_F$ and $\boldsymbol{d}_M$ are functions of elastic accelerations and cross couplings between rigid-body and elastic velocities (Coriolis effects)~\cite{Wang2019c,Meirovitch2003}. If $\boldsymbol{d}_F$ and $\boldsymbol{d}_M$ are set to zero, while ${\boldsymbol{S}}$ and ${\boldsymbol{J}}$ are kept constant, then the rigid-body dynamics are retrieved. If in the resulting simplified dynamics, further set ${\boldsymbol{S}}$ to zero, then the conventional rigid-body dynamics where $O_b$ is coincide with aircraft c.m. are retrieved. $\boldsymbol{d}_F$, $\boldsymbol{d}_M$ and the dependencies of ${\boldsymbol{S}}$, ${\boldsymbol{J}}$ on elastic states essentially reflect the inertial couplings between rigid-body and elastic motions. 

As compared to the inertial couplings, aerodynamic couplings normally have more dominant effects. This is also the only coupling aspect considered in the mean-axes method~\cite{WASZAK1988}. As discussed in subsection~\ref{subsection_ase_model}, rigid-body motions excite wing aeroelastic dynamics through the input $\boldsymbol{\alpha}_{qs,r}$. On the other hand, $\boldsymbol{F}_{\text{tot}}^B$ and $\boldsymbol{M}_{\text{tot}}^B$ are contributed by $\boldsymbol{f}_{\text{aero}}$ (Eq.~\eqref{eq_aero}), which is a function of aeroelastic states. It is noteworthy that drag is not included in $\boldsymbol{f}_{\text{aero}}$, while it should be considered in $\boldsymbol{F}_{\text{tot}}^B$ and $\boldsymbol{M}_{\text{tot}}^B$. For each aerodynamic strip, the local drag coefficient is calculated using the quadratic expression $C_D = C_{D_0} + k_D C_L^2$. Since the fuselage and tails are assumed to be rigid, the quasi-steady strip theory is used for calculating the distributed lift and drag on them. Denote the distance vector from $O_b$ to an arbitrary strip as $\boldsymbol{r}_i$ (expressed in $\mathcal{F}_B$), then the local airspeed is derived as $\boldsymbol{V}_{a_i} = \boldsymbol{V}_b + \Tilde{\boldsymbol{\omega}}_b \boldsymbol{r}_i -  \boldsymbol{C}_{BI}[0,0,  w_g(t)]^\mathsf{T}$. During simulations, the turbulence velocity $w_g(t)$ is calculated by interpolating the spatial vertical atmospheric disturbance field at the local strip position $\boldsymbol{R}_b +\boldsymbol{C}_{BI}^\mathsf{T}  \boldsymbol{r}_i$~\cite{Wang2019c}. From $\boldsymbol{V}_{a_i}$, the local dynamic pressure and aerodynamic angles can be calculated. 


\section{Control Design}
\label{sec_control}
After presenting the flight dynamics, kinematics, and wing aeroservoelastic dynamics, the flexible aircraft trajectory tracking control architecture will be shown in this section. Cascaded control loops are widely adopted in flexible aircraft control~\cite{Qi2019b,Zhaoxiaowei}. The control architecture proposed in this paper contains four cascaded control loops: position control, flight path control, attitude control, and optimal multi-objective wing control. These control loops will be designed in the following subsections:

\subsection{Position Control}
The kinematics of aircraft positions are given in Eq.~\eqref{eq_position}. The control objective is to make the aircraft follow a pre-designed three-dimensional trajectory $[X^\text{ref},Y^\text{ref},Z^\text{ref}]^\mathsf{T}$. Since there is no model uncertainty in the position kinematics, the nonlinear dynamic inversion control is adopted. Design the virtual control for the lateral and vertical positions as 
\begin{equation}
    \nu_Y = \dot Y^\text{ref} + K_Y (Y^\text{ref} -Y),~~\nu_Z = \dot Z^\text{ref} + K_Z (Z^\text{ref} -Z)
\end{equation}
where $ K_Y$ and $ K_Z$ are positive proportional gains, then when $\dot Y= \nu_Y$, and $\dot Z = \nu_Z$, the lateral and vertical tracking errors converge to zero exponentially. Replace $\dot Y$ and $\dot Z$ in Eq.~\eqref{eq_position} by their virtual controls, and then invert the resulting equations, the references for the kinematic azimuth angle and the flight path angles are obtained as
\begin{equation}
    \chi^\text{ref} = \arcsin({\nu_Y}/(V \cos \gamma)),~~\gamma^\text{ref} = -\arcsin({\nu_Z}/{V})
    \label{ref_chi_gamma}
\end{equation}

\subsection{Flight Path Control}
\label{subsection_path}
The control objective of this loop it to make the flight path angles track their references designed in Eq.~\eqref{ref_chi_gamma}. Because the total external force vector contains aerodynamic, gravitational and propulsion forces, the aircraft translational dynamics given in Eq.~\eqref{eq_Vxigamma} can be rewritten as
\begin{equation}
    \begin{bmatrix}
    \dot V \\
    \dot \chi \\
    \dot \gamma
    \end{bmatrix} = 
    \begin{bmatrix}
    1 & 0&0\\
    0& \frac{1}{V \cos\gamma} & 0 \\
    0&0& -\frac{1}{V}
    \end{bmatrix} 
    \left(
    \frac{\boldsymbol{C}_{VB}}{m} 
    \begin{bmatrix}
    T \\ 0\\ 0
    \end{bmatrix}
    +  \frac{\boldsymbol{C}_{VI}}{m} 
    \begin{bmatrix}
    0 \\ 0\\ mg
    \end{bmatrix} 
    +  \frac{\boldsymbol{C}_{VA}}{m} 
    \begin{bmatrix}
    -D \\ C\\ -L
    \end{bmatrix}+ 
    \begin{bmatrix}
    d_V \\ d_\chi\\ d_\gamma
    \end{bmatrix}
    \right)
    \label{eq_Vxig}
\end{equation}
in which $[T,0,0]^\mathsf{T}$ is the thrust vector. $D$, $C$ and $L$ are the total drag, side force, and lift of the aircraft, which are defined in the aerodynamic axes $\mathcal{F}_A$. $[d_V,d_\chi,d_\gamma]^\mathsf{T}$ includes $\boldsymbol{d}_F$ in Eq.~\eqref{eq_Vxigamma} and the cross-coupling terms caused by the non-zero first moment of area. Substituting the expressions of the direction cosine matrices into Eq.~\eqref{eq_Vxig}, the dynamics of the flight path angle are
\begin{equation}
    \dot \gamma = -\frac{1}{mV} ( -T \sin \alpha \cos \mu + mg \cos\gamma - L \cos \mu +d_\gamma )
    \label{eq_gamma}
\end{equation}

From the physical point of view, one of the most effective ways of changing aircraft flight path is by changing the total lift, which is dominated by the angle of attack $\alpha$. Therefore, $\alpha$ is selected as a control input to the dynamics of $\gamma$. Denote the flight path angle dynamics in Eq.~\eqref{eq_gamma} as $\gamma = f_\gamma (\boldsymbol x, u)+d_\gamma$, where the vector $\boldsymbol x$ include the rigid-body and elastic states except $\alpha$. It can be seen from Eq.~\eqref{eq_gamma} that $f_\gamma(\boldsymbol x, u)+d_\gamma$ is non-affine in control. Conventional nonlinear control methods including feedback linearization and backstepping cannot be directly applied. In this paper, the novel incremental sliding mode control method is adopted, which not only applies to non-affine in control dynamic systems, but also has enhanced robustness to model uncertainties~\cite{Wang2018b}. 

Design the sliding surface as $\sigma_\gamma = \gamma -\gamma^\text{ref}$. The control objective is to design a reference for $\alpha$, such that when $\alpha$ tracks its reference through inner-loop control, $\sigma_\gamma $ also converges to zero. Denote the sampling interval as $\Delta t$, the incremental dynamic equation is derived by taking the first-order Taylor series expansion of Eq.~(\ref{eq_gamma}) around the condition at $t-\Delta t$ (denoted by the subscript 0) as
\begin{equation}
 \dot \gamma = f_{\gamma} (\boldsymbol x, u)+ d_\gamma = \dot \gamma_0+ \frac{\partial f_{\gamma}}{\partial u} \bigg|_0 \Delta u + \frac{\partial f_{\gamma}}{\partial \boldsymbol{x}} \bigg|_0 \Delta \boldsymbol{x} + \Delta d_\gamma + {R}_1
 \label{incre_dynamic}
\end{equation}
in which $\Delta \boldsymbol x$ and $\Delta u$ respectively represents the variations of $\boldsymbol x$ and $u$ in one sampling time step $\Delta t $. $\Delta d_\gamma$ is the variations of $ d_\gamma$ in $\Delta t$. ${R}_1$ is the expansion remainder, whose Lagrange form is
\begin{equation}
{R}_1 = \frac{1}{2}  \frac{\partial^2 f_{\gamma}}{\partial^2 \boldsymbol x}\bigg|_{m} \Delta \boldsymbol x^2 + \frac{\partial^2 f_{\gamma}}{\partial \boldsymbol x \partial  u}\bigg|_{m} \Delta \boldsymbol x \Delta  u  + \frac{1}{2}\frac{\partial^2 f_{\gamma}}{\partial^2  u}\bigg|_{m} \Delta u^2
\label{R_1}
\end{equation}
in which $(\cdot)|_m$ means evaluating $(\cdot)$ at a condition where $\boldsymbol{x} \in (\boldsymbol{x}(t-\Delta t),\boldsymbol{x}(t))$, ${u} \in ({u}(t-\Delta t),{u}(t))$. The incremental sliding mode control law for stabilizing $\sigma_\gamma$ is then designed as
\begin{equation}
\Delta u = (\nu_n+\nu_s+\dot{ \gamma}^\text{ref} - \dot \gamma_0)/\bar{G}_0
\label{u_gamma}
\end{equation}
where the virtual control $\nu_n$ is designed to stabilize the nominal sliding dynamics, while the virtual control $\nu_s$ is a robustify virtual control for disturbance rejection. The specific expressions of these two virtual control terms will be presented later. The real control input equals $\Delta u + u_0$, where $u_0$ denotes the control input at $t-\Delta t$. In Eq.~\eqref{u_gamma}, $\bar{G}_0$ represents the estimation of the control effectiveness $\frac{\partial f_{\gamma}}{\partial u} \big|_0$. It can be seen from Eq.~\eqref{eq_gamma} that 
\begin{equation}
    G_0 = \frac{\partial f_{\gamma}}{\partial u} \bigg|_0 = \frac{\cos \mu}{mV} \left( T\cos \alpha + \frac{\partial L}{\partial \alpha}\right) \bigg|_0
\end{equation}
its estimation is calculated as 
\begin{equation}
    \bar{G}_0 =  \frac{\cos \mu}{mV} \left( T\cos \alpha + q_{\infty} S_w \bar{C}_{L\alpha}  \right) \big|_0
    \label{eq_G_0}
\end{equation}
where $q_{\infty}$ is the dynamic pressure and $S_w$ is the wing area. Substituting Eq.~\eqref{u_gamma} into Eq.~\eqref{incre_dynamic}, then the closed-loop dynamics are 
\begin{equation}
    \dot {\sigma}_\gamma = \dot {\gamma} -\dot {\gamma}^\text{ref} 
     = \dot \gamma_0+ (G_0 /\bar{G}_0) (\nu_n+\nu_s+\dot{ \gamma}^\text{ref} - \dot \gamma_0) - \dot {\gamma}^\text{ref} +\delta(\boldsymbol{x},\Delta t) + \Delta d_\gamma
     \triangleq
    \nu_n+\nu_s+  \varepsilon_\gamma
    \label{closed-loop}
\end{equation}
where $\delta(\boldsymbol{x},t)$ is the closed-loop value of the state variations and the expansion reminder:
\begin{equation}
    \delta(\boldsymbol{x},t) = \left[\frac{\partial f_{\gamma}}{\partial \boldsymbol{x}} \bigg|_0 \Delta \boldsymbol{x} + {R}_1 \right]_{u=u_0+\Delta u}
    \label{eq_delta}
\end{equation}

$\varepsilon_\gamma$ in Eq.~\eqref{closed-loop} is the lumped perturbation term, which is expressed as
\begin{equation}
    \varepsilon_\gamma = \delta(\boldsymbol{x},\Delta  t) + \Delta d_\gamma + (G_0 /\bar{G}_0-1) (\nu_n+\nu_s+\dot{ \gamma}^\text{ref} - \dot \gamma_0) 
    \label{epsilon_gamma}
\end{equation}

\begin{assumption}
\label{ass_1}
  $\delta(\boldsymbol{x},\Delta  t)$ in Eq.~\eqref{eq_delta} and $\Delta d_\gamma$ in Eq.~\eqref{incre_dynamic} satisfy $|\delta(\boldsymbol{x},\Delta  t)|<\bar{\delta}$, $|\Delta d_\gamma|<\bar{d}_\gamma$. 
\end{assumption}

Since $\boldsymbol{x}$ is continuously differentiable, then $\lim_{\Delta t \rightarrow 0}  \|  \Delta \boldsymbol x  \| = 0$. If the partial derivatives of $f_\gamma$ with respect to $\boldsymbol{x}$, up to any order, are bounded, then in view of Eqs.~\eqref{R_1} and~\eqref{eq_delta}, the absolute value of $\delta(\boldsymbol{x},\Delta  t)$ approaches zero as $\Delta t$ decreases. Therefore, this assumption holds when $\Delta t$ is sufficiently small.

\begin{proposition}
Under Assumption~\ref{ass_1}, if $~~0< G_0/\bar{G}_0<2$, then when the sampling frequency $f_s$ is sufficient high, the uncertainty term $\varepsilon_\gamma$ in Eq.~\eqref{epsilon_gamma} is ultimately bounded.
\end{proposition}

\noindent\textbf{\textit{Proof:}}
Substituting Eq.~\eqref{u_gamma} into Eq.~\eqref{incre_dynamic}, the resulting closed-loop dynamics are $\dot \gamma = \nu_n+\nu_s +\dot{ \gamma}^\text{ref} + \varepsilon_\gamma$. This closed-loop dynamics are also valid at $t-\Delta t$, thus $\dot \gamma_0 = (\nu_n+\nu_s+ \dot{ \gamma}^\text{ref})|_0 + \varepsilon_{\gamma,0}$. Therefore, $\varepsilon_\gamma$ in Eq.~\eqref{epsilon_gamma} can be rewritten as  
\begin{eqnarray}
    \varepsilon_\gamma &=& \delta(\boldsymbol{x},\Delta  t) + \Delta d_\gamma + (G_0 /\bar{G}_0-1) (\nu_n+\nu_s+\dot{ \gamma}^\text{ref} - ((\nu_n+\nu_s+\dot{ \gamma}^\text{ref})|_0 + \varepsilon_{\gamma,0}))  \nonumber \\
    &=& (1-G_0 /\bar{G}_0)\varepsilon_{\gamma,0} - (1-G_0 /\bar{G}_0) (\nu_\gamma - \nu_{\gamma,0}) + \delta(\boldsymbol{x},\Delta  t) + \Delta d_\gamma
\end{eqnarray}
where $\nu_\gamma = \nu_n+\nu_s+\dot{ \gamma}^\text{ref}$. These virtual control terms are all continuous in time, thus under sufficiently high sampling frequency, the variation of $\nu$ in $\Delta t$ has an upper bound, i.e., $|\nu_\gamma - \nu_{\gamma,0}|<\bar{\nu}_\gamma$. Analogous to the proof of Theorem~1 in~\cite{Wang2018b}, the ultimate boundedness of $\varepsilon_\gamma$ can be proved.
\hfill$\square$

The boundedness of $\varepsilon_\gamma$ makes it feasible to compensate for its influences using robust control. In this paper, $\nu_s$ in Eq.~\eqref{u_gamma} is designed using a super-twisting sliding mode observer. Design an auxiliary sliding variable as $s_\gamma = \sigma_\gamma -\int \nu_n ~\text{d}t$, then using Eq.~\eqref{closed-loop}, the dynamics of $s_\gamma$ are 
\begin{equation}
    \dot {s}_\gamma = \dot {\sigma}_\gamma - \nu_n = \nu_s+  \varepsilon_\gamma
\end{equation}

Further design the observer virtual control as 
\begin{equation}
    \nu_s = -\lambda_\gamma |s|^{0.5}\text{sign}(s) -\beta_\gamma \int \text{sign} (s)~ \text{d}t
\end{equation}

Denote the upper bound of $\varepsilon_\gamma$ as $\bar{\varepsilon}_\gamma$, and design $\lambda_\gamma = 1.5 \bar{\varepsilon}_\gamma^{0.5}$, $\beta = 1.1\bar{\varepsilon}_\gamma$, then $s=\dot s = 0$ is established in finite-time~\cite{Wang2019t}. On this sliding surface, $\nu_s$ provides a real-time observation of the uncertainty term $-\varepsilon_\gamma$. Moreover, the nominal dynamics $\dot {\sigma}_\gamma = \nu_n$ are retrieved on the sliding surface regardless of uncertainties. Simply design the nominal virtual control as $\nu_n = - K_\sigma {\sigma}_\gamma$, with $K_\sigma>0$, then ${\sigma}_\gamma$ converges to zero exponentially. The observer gains can also be made adaptive,
which removes the per-knowledge requirement on the uncertainty bound~\cite{Edwards2016}.

\begin{remark}
\rm The only model information required by the incremental sliding mode control in Eq.~\eqref{u_gamma} is the control effectiveness $G_0$. In view of Eq.~\eqref{eq_G_0}, the model parameters in $G_0$ are only the wing area $S_w$ and the aircraft lift-slope $C_{L\alpha}$. These two parameters are easily known from aircraft overall design.
\end{remark}

\begin{remark}
\rm The presented flight path control differs from the one in~\cite{Lu2016} in three aspects: 1) the residual uncertainty term $\varepsilon_\gamma$ is observed and is compensated for in this paper. 2) In Eq.~\eqref{eq_Vxig}, the aerodynamic forces are expressed in the aerodynamic axes, while in~\cite{Lu2016}, they are expressed in the body axes. Consequently, the aerodynamic coefficient used in the proposed controller is only $C_{L\alpha}$, while Ref.~\cite{Lu2016} requires both $C_{X\alpha}$ and $C_{Z\alpha}$ ($X,Z$ are the aerodynamic forces expressed in the body-fixed frame). As compared to $C_{X\alpha}$ and $C_{Z\alpha}$, the state-dependency of $C_{L\alpha}$ is lower, which simplifies its identification process. 3) In Ref.~\cite{Lu2016}, throttle control is also included in the incremental control loop. However, aircraft throttle normally has much lower bandwidth than the aerodynamic control surfaces. In view of this, a separate throttle controller ($\delta_T$) for maintaining airspeed is adopted in this paper. 
\end{remark}

After presenting the flight path angle ($\gamma$) control law, the reference tracking problem for the kinematic azimuth angle $\chi$ will be solved. Recall Eq.~\eqref{eq_Vxig}, the dynamics of $\chi$ are
\begin{equation}
    \dot \chi  = \frac{1}{V\cos\gamma} (T \sin \alpha \cos \mu + L \sin \mu +d_\chi )
    \label{eq_chi}
\end{equation}

In view of these dynamics, the kinematic bank angle $\mu$ is chosen as the control input. Replace $\dot \gamma$ in Eq.~\eqref{eq_gamma} and $\dot \chi $ in Eq.~\eqref{eq_chi} by their virtual control $\nu_\gamma$ and $\nu_\chi$ respectively, and then invert the resulting dynamics, the reference for $\mu$ is designed as
\begin{equation}
    \mu^\text{ref} = \arctan \left( \frac{\nu_\chi V \cos \gamma }{\nu_\gamma V+g\cos \gamma} \right)
\end{equation}
in which $\nu_\gamma$ has been designed in the preceding texts, while the virtual control $\nu_\chi$ is designed as $\nu_\chi = \dot \chi^{\text{ref}} + K_\chi (\chi^{\text{ref}} - \chi)$ with $K_\chi>0$. 

\subsection{Attitude Control}
\label{sub_set_attitude}
The flexible aircraft attitude kinematics are given in Eq.~\eqref{kine_atti}, while the dynamics of angular rates are given in Eq.~\eqref{eq_dot_pqr}. The control objective of this attitude control loop is to make $\mu,\alpha$ and $\beta$ track their references. $\mu^{\text{ref}}$ and $\alpha^{\text{ref}}$ have been designed in subsection~\ref{subsection_path}, while the $\beta^{\text{ref}}$ is set to zero for mitigating side force. From a physical point of view, the attitude of an aircraft can be changed by creating control moments around $O_bx_b$, $O_by_b$ and $O_bz_b$ axes. For the aircraft configuration shown in Fig.~\ref{fig:ref_frames}, the pitch and yaw control moments can be generated by the deflections of elevator $\delta_e$ and rudder $\delta_r$. On the other hand, the difference between the left and right wing root bending moments is essentially the main source of aircraft rolling moment. Therefore, $\delta_e$, $\delta_r$ and the wing root bending moment difference $M_{\phi,\text{diff}} = M_{\phi,l} - M_{\phi,r}$ are chosen as control input variables in this loop. The wing root bending moment can be measured by strain gauges at a sampling frequency around 1~kHz. 

Define the state vectors as $\boldsymbol{x}_1 =[\mu,\alpha,\beta]^\mathsf{T}$, $\boldsymbol{x}_2 =[p,q,r]^\mathsf{T}$, and define the control vector as $\boldsymbol{u} = [\delta_e,\delta_r,M_{\phi,\text{diff}}]^\mathsf{T}$, then the attitude kinematics and dynamics in Eqs.~\eqref{kine_atti} and~\eqref{eq_dot_pqr} can be represented as
\begin{eqnarray}
    \dot {\boldsymbol x}_1 &=& \boldsymbol f_1(\boldsymbol x_1) +\boldsymbol G_1(\boldsymbol x_1) \boldsymbol x_2 \nonumber \\
\dot {\boldsymbol x}_2 &=& \boldsymbol f_2(\boldsymbol x_1,\boldsymbol x_2,\boldsymbol{x}_e) +\boldsymbol G_2(\boldsymbol x_1,\boldsymbol x_2,\boldsymbol{x}_e,\boldsymbol{u}) + \boldsymbol{d} 
\end{eqnarray}
where $\boldsymbol{d} = \boldsymbol{J}^{-1} \boldsymbol{d}_M$, while $\boldsymbol{x}_e = [\dot {\boldsymbol{x}}_s,\boldsymbol{x}_s,\boldsymbol{z}_a]^\mathsf{T}$ represents the state vector in Eq.~\eqref{eq_wing_aero}. Choose the controlled output as $\boldsymbol{ y} = {\boldsymbol x}_1$, then the control objective is to make $\boldsymbol{ y}$ track its reference $\boldsymbol{ y}^{\text{ref}}$. In view of the kinematics in Eq.~\eqref{kine_atti}, there is no model uncertainty in $\boldsymbol f_1$ and $\boldsymbol G_1$. On the contrary, $\boldsymbol f_2$ and $\boldsymbol G_2$ contains model uncertainties, and are functions of elastic states. In addition, $\boldsymbol G_2$ in non-affine in control, which makes direct applications of feedback linearization and backstepping infeasible.  

The novel incremental backstepping sliding mode control proposed in~\cite{Wang2019} is adopted to handle this output tracking problem. Define the error variable as $\boldsymbol{z}_1 = \boldsymbol{y} - \boldsymbol{ y}^{\text{ref}}$, then 
\begin{equation}
    \dot{\boldsymbol z}_1 = \boldsymbol f_1 + \boldsymbol G_1 \boldsymbol x_2 - \dot{\boldsymbol{ y}}^{\text{ref}}
\end{equation}

Consider a candidate Lyapunov function $V_1(\boldsymbol z_1) = \frac{1}{2}\boldsymbol z_1^\mathsf{T}  \boldsymbol z_1$. In order to make $\dot{V}_1(\boldsymbol z_1) \leq 0$, the reference for $\boldsymbol{x}_2$ is designed as
\begin{equation}
   \boldsymbol{x}_2^{\text{ref}} =  \boldsymbol G_1^{-1} (-\boldsymbol f_1 - \boldsymbol K_1 \boldsymbol z_1 +  \dot{\boldsymbol y}^{\text{ref}} ) 
   \label{x_2}
\end{equation}
in which $\boldsymbol K_1$ is a positive definite diagonal gain matrix. Further define the tracking error for $\boldsymbol{x}_2$ as $\boldsymbol z_2 = \boldsymbol x_2 - \boldsymbol x_{2}^{\text{ref}}$, then the resulting dynamics are $\dot{ \boldsymbol z}_2 =\boldsymbol f_2(\boldsymbol x_1,\boldsymbol x_2,\boldsymbol{x}_e) +\boldsymbol G_2(\boldsymbol x_1,\boldsymbol x_2,\boldsymbol{x}_e,\boldsymbol{u}) + \boldsymbol{d}  - \dot {\boldsymbol{x}}_2^{\text{ref}}$. Denote the augmented state vector as $\boldsymbol{x} = [\boldsymbol{x}_1^\mathsf{T},\boldsymbol{x}_2^\mathsf{T},\boldsymbol{x}_e^\mathsf{T}]^\mathsf{T}$. The incremental dynamics of $\boldsymbol{x}_2$ are derived by taken the first-order Taylor series expansion around the condition at $t-\Delta t$ (denoted by the subscript 0) as:
\begin{eqnarray}
\dot{\boldsymbol x}_2 &=&  \dot{\boldsymbol x}_{2,0} 
+\frac{\partial [ \boldsymbol f_2(\boldsymbol x) + \boldsymbol G_2(\boldsymbol x,\boldsymbol{u})]}{\partial \boldsymbol u}   \bigg|_0  \Delta \boldsymbol u
+\frac{\partial [ \boldsymbol f_2(\boldsymbol x) + \boldsymbol G_2(\boldsymbol x,\boldsymbol{u})]}{\partial \boldsymbol x}   \bigg|_0  \Delta \boldsymbol x 
+\Delta \boldsymbol d + \boldsymbol R_1'
\label{eq_incre}
\end{eqnarray}
where $\Delta \boldsymbol x = \boldsymbol x - \boldsymbol x_0$, $\Delta \boldsymbol u = \boldsymbol u - \boldsymbol u_0$, and $\Delta \boldsymbol d = \boldsymbol d - \boldsymbol d_0$ respectively denotes the variations of states, control inputs, and disturbances, in one incremental time step. $\boldsymbol{R}_1'$ in Eq.~(\ref{eq_incre}) is the expansion remainder, whose Lagrange form is
\begin{equation}
\boldsymbol{R}_1' = \frac{1}{2}  \frac{\partial^2 [ \boldsymbol f_2(\boldsymbol x) + \boldsymbol G_2(\boldsymbol x,\boldsymbol{u})]}{\partial^2 \boldsymbol x}\bigg|_{m} \Delta \boldsymbol x^2 + \frac{\partial^2 [ \boldsymbol f_2(\boldsymbol x) + \boldsymbol G_2(\boldsymbol x,\boldsymbol{u})]}{\partial  \boldsymbol x \partial  \boldsymbol u}\bigg|_{m} \Delta \boldsymbol x \Delta \boldsymbol u  + \frac{1}{2} \frac{\partial^2 [ \boldsymbol f_2(\boldsymbol x) + \boldsymbol G_2(\boldsymbol x,\boldsymbol{u})]}{\partial^2 \boldsymbol u }\bigg|_{m} \Delta \boldsymbol u^2 
\label{R_11}
\end{equation}
in which $(\cdot)|_m$ means evaluating $(\cdot)$ at a condition where $\boldsymbol{x} \in (\boldsymbol{x}(t-\Delta t),\boldsymbol{x}(t))$, $\boldsymbol{u} \in (\boldsymbol{u}(t-\Delta t),\boldsymbol{u}(t))$, $\boldsymbol{d} \in (\boldsymbol{d}(t-\Delta t),\boldsymbol{d}(t))$. To stabilize $\boldsymbol{z}_2$, the incremental backstepping sliding mode control input is designed as 
\begin{equation}
   \Delta \boldsymbol{u}_{\text{ibsmc}} = \bar{\boldsymbol G}_{2}^{-1} (\boldsymbol \nu_c +\boldsymbol \nu_s -\dot{\boldsymbol x}_{2,0})
   \label{eq_u_ibsmc}
\end{equation}
where $\bar{\boldsymbol G}_{2}$ is the estimation of the control effectiveness matrix ${\boldsymbol G}_{2} = \frac{\partial [ \boldsymbol f_2(\boldsymbol x) + \boldsymbol G_2(\boldsymbol x,\boldsymbol{u})]}{\partial \boldsymbol u}   \big|_0 $. The virtual control $\boldsymbol \nu_c = - \boldsymbol K_{2} \boldsymbol z_{2} + \dot {\boldsymbol{x}}_2^{\text{ref}} - \boldsymbol G_{1}^\mathsf{T} \boldsymbol z_{1}$ is used for stabilizing the nominal dynamics. $\boldsymbol K_{2}$ is a positive definite diagonal gain matrix. Consider the sliding surface $\boldsymbol \sigma = \boldsymbol z_2 = \boldsymbol 0$, the robustify virtual control $\boldsymbol \nu_s$ is designed as:
\begin{equation}
\boldsymbol \nu_s = - \boldsymbol K_s \text{sig}({\boldsymbol \sigma})^{\boldsymbol \gamma_s} = - [K_{s,1}|\sigma_1|^{\gamma_{s,1}}\text{sign}(\sigma_1), K_{s,2}|\sigma_2|^{\gamma_{s,2}}\text{sign}(\sigma_2),K_{s,3}|\sigma_3|^{\gamma_{s,3}}\text{sign}(\sigma_3)]^\mathsf{T}
\label{nu_s}
\end{equation}
where $K_{s,i} >0,~\gamma_{s,i} \in(0,1)$. In the conventional sliding mode control, the discontinuous sign function has to be smoothened for chattering reduction. This process is not needed here, since without any approximation, $|\sigma_i|^{\gamma_i}\text{sign}(\sigma_i)$ itself is a continuous function of $\sigma_i$. Substituting Eq.~\eqref{eq_u_ibsmc} into Eq.~\eqref{eq_incre}, the closed-loop dynamics are 
\begin{equation}
    \dot {\boldsymbol x }_2 = \boldsymbol \nu_c + \boldsymbol{ \delta}(\boldsymbol{x},\Delta t) + (\boldsymbol G_2 \bar{\boldsymbol G}_2^{-1} - \boldsymbol I)  (\boldsymbol \nu_c - \dot{\boldsymbol x}_{2,0} ) + \boldsymbol G_2 \bar{\boldsymbol G}_2^{-1}\boldsymbol \nu_s+   \Delta \boldsymbol d  \triangleq \boldsymbol \nu_c +  \boldsymbol G_2 \bar{\boldsymbol G}_2^{-1} \boldsymbol  \nu_s + \boldsymbol {\varepsilon}_{\text{ibs}}
    \label{eq_V_n_ibs}
\end{equation}
where $\boldsymbol{ \delta}(\boldsymbol{x},\Delta t) $ equals the summation of the closed-loop values of $\boldsymbol{R}_1'$ (Eq.~\eqref{R_11}) and $\frac{\partial [ \boldsymbol f_2(\boldsymbol x) + \boldsymbol G_2(\boldsymbol x,\boldsymbol{u})]}{\partial \boldsymbol x}   \big|_0  \Delta \boldsymbol x $ (Eq.~\eqref{eq_incre}). 

\begin{proposition}
If $\|\boldsymbol I-  \boldsymbol G_2  \bar {\boldsymbol G}_2^{-1}  \| \leq \bar b <1$ for all $t$, and if $\|\boldsymbol \delta( \boldsymbol x,\Delta t) \|\leq \bar{\delta}$, $\| \Delta \boldsymbol d\|\leq  \overline{\Delta d}$, then under sufficiently high sampling frequency, $\boldsymbol \varepsilon_{\text{ibs}}$ given by Eq.~(\ref{eq_V_n_ibs}) is bounded for all $t$, and is ultimately bounded by $\frac{\overline{\Delta \nu}_c \bar{ b} + \bar{\delta} +\overline{\Delta d}}{1-\bar{b}}$, where $\overline{\Delta \nu}_c$ is the upper bounds of $\Delta \boldsymbol \nu_c$.
\end{proposition}

This proposition can be proved following the proofs of Theorem~1 in~\cite{Wang2019}. It puts requirement on the control effectiveness estimation, i.e., $\boldsymbol G_2  \bar {\boldsymbol G}_2^{-1}$ has to be diagonally dominated. 

\begin{proposition}
If $\boldsymbol \varepsilon_{\text{ibs}}$ (Eq.~(\ref{eq_V_n_ibs})) is bounded, then using the control incremental in Eq.~\eqref{eq_u_ibsmc}, both $\boldsymbol{z}_1$ and $\boldsymbol{z}_2$ are ultimately bounded.  
\end{proposition}

\noindent\textbf{\textit{Proof:}}
Consider a candidate Lyapunov function $V_2(\boldsymbol z_1,\boldsymbol z_2) = \frac{1}{2}(\boldsymbol z_1^\mathsf{T}  \boldsymbol z_1 + \boldsymbol z_2^\mathsf{T}  \boldsymbol z_2 )$, using Eqs.~\eqref{x_2} and~\eqref{eq_u_ibsmc}, the time derivative of $V_2$ are derived as
\begin{eqnarray}
\dot V_2 &=&  -\boldsymbol z_1^\mathsf{T} \boldsymbol K_1 \boldsymbol z_1  -\boldsymbol z_2^\mathsf{T} \boldsymbol K_2 \boldsymbol z_2 + \boldsymbol z_2^\mathsf{T}( \boldsymbol \varepsilon_{\text{ibs}} - \boldsymbol G_n \bar{\boldsymbol G}_n^{-1} \boldsymbol K_s \text{sig}({\boldsymbol \sigma})^{\boldsymbol \gamma}) \nonumber \\ &\leq& -\boldsymbol z_1^\mathsf{T} \boldsymbol K_1 \boldsymbol z_1  -\boldsymbol z_2^\mathsf{T} \boldsymbol K_2 \boldsymbol z_2 +  \sum_{i=1}^{3} (|\sigma_i| |\varepsilon_{\text{ibs},i}| + \bar{b} K_{s,i} |\sigma_i|^{\gamma_{s,i+1}} - K_{s,i}|\sigma_i|^{\gamma_{s,i+1}} ) \nonumber \\
&\leq& -\boldsymbol z_1^\mathsf{T} \boldsymbol K_1 \boldsymbol z_1  -\boldsymbol z_2^\mathsf{T} \boldsymbol K_2 \boldsymbol z_2 -  \sum_{i=1}^{3}\rho_i |\sigma_i|,
~~~~~\forall |\sigma_{i}|\geq \bigg(\frac{\rho_i + |\varepsilon_{\text{ibs},i}|}{(1-\bar{b}) K_{s,i}} \bigg)^{\frac{1}{\gamma_{s,i}}},~~~\forall \rho_i >0
\end{eqnarray}

The above equation proves that the ultimate bound~\cite{Khalil} of $\sigma_i$ equals $\left(\frac{\rho_i + |\varepsilon_{\text{ibs},i}|}{(1-\bar{b}) K_{s,i}} \right)^{\frac{1}{\gamma_{s,i}}}$, whose size can be made arbitrarily small by increasing $K_{s,i} $ and reducing $\gamma_{s,i}$. Because $\boldsymbol \sigma = \boldsymbol z_2$, $\boldsymbol{K}_1 $ and $\boldsymbol{K}_2 $ are positive definite, then both $\boldsymbol{z}_1$ and $\boldsymbol{z}_2$ are ultimately bounded. \hfill$\square$

\begin{remark}
\rm Incremental backstepping sliding mode control has lower model dependency than feedback linearization and backstepping sliding mode control. The only model information needed for implementation is the control effectiveness $\bar {\boldsymbol{G}}_2$. Even so, by virtue of its sensor-based nature, incremental backstepping sliding mode control actually has enhanced robustness against model uncertainties, sudden actuator faults and structural damage~\cite{Wang2019,Wang2019t}. Under the same perturbation circumstance, there exists a sampling frequency such that the bound of $\boldsymbol {\varepsilon}_{\text{ibs}}$ is smaller than that of the residual error under backstepping control. This property enables incremental backstepping sliding mode control to passively resist a wider range of perturbations with much lower sliding mode control gains~\cite{Wang2019,Wang2019t}.
\end{remark}

\subsection{Optimal Multi-objective Wing Control}
\label{sub_sec_optimal}
In subsection~\ref{sub_set_attitude}, the reference values for $ \delta_e,\delta_r,M_{\phi,\text{diff}}$ have been designed for aircraft attitude control purposes. The references for $ \delta_e $ and $\delta_r$ can be directly given to the actuators of elevator and rudder, whereas $M_{\phi,\text{diff}}^{\text{ref}}$ still needs to be achieved by aircraft trailing-edge control surfaces. The aircraft model considered in this paper has fourteen distributed flaps, which makes the achievement of $M_{\phi,\text{diff}}^{\text{ref}}$ an \textit{over-actuated} problem. As a consequence, after $M_{\phi,\text{diff}}^{\text{ref}}$ is achieved for the purpose of bank angle tracking, the wing system still have \textit{control redundancy} to achieve other objectives such as maneuver load alleviation, gust load alleviation, flutter suppression, etc. In view of this, an optimal multi-objective wing control law will be designed in this subsection. 

Wing-root shear force ${F}_w$ and wing-root bending moment ${M}_\phi$ are two important wing load indicators. Theirs values would deviate from the trimmed condition during aircraft maneuvers, under the excitations of atmospheric disturbances, as well as under flap inputs. In a trimmed flight, when an aircraft encounters disturbances, a reasonable control objective is to maintain ${F}_w$ and ${M}_\phi$ at their trimmed values. However, this objective is not applicable during aircraft maneuvers for two reasons: 1) in view of Eq.~\eqref{eq_gamma}, the wing-root shear force ${F}_w$, which is strongly coupled with the total lift on a half-wing, is the main medium for changing flight path angle $\gamma$; 2) as discussed in subsection~\ref{sub_set_attitude}, the difference between the left and right wing-root bending moments $M_{\phi,\text{diff}}$ is the main medium for aircraft roll control. 

Based on these discussions, it is essential to identify the loads that are \textit{necessary} to achieve commanded maneuvers, while reducing any \textit{excessive} load induced by either maneuvers or external disturbances. A novel reference generator is designed to fulfill this goal. The designed references for the left and right wing-root shear forces are
\begin{equation}
    F_{w,r}^{\text{ref}} = F_{w,l}^{\text{ref}} = F_{w}^{\text{trim}} + G_{F_w}(s) (\alpha^{\text{ref}} - \alpha^{\text{trim}})
    \label{F_w_ref}
\end{equation}

In Eq.~\eqref{F_w_ref}, $\alpha^{\text{ref}}$ is the angle of attack reference designed in subsection~\ref{subsection_path}. $G_{F_w}(s)$ represents an estimated mapping from $\alpha$ input to the open-loop shear force response of a half wing. In this paper, $G_{F_w}(s)$ is designed as a low-pass filer, with $s$ represents the Laplace variable. The zero-frequency gain of $G_{F_w}(s)$ equals $q_{\infty} S_w {C}_{L\alpha_w}/2 $, where $S_w$ is the wing area and ${C}_{L\alpha_w}$ is the lift-slope of the wing. The dynamics of $G_{F_w}(s)$ are caused by the time-dependent circulatory effects. Therefore, the time-constant of $G_{F_w}(s)$ can be either calculated using Eq.~\eqref{eq_state_space}, or be identified online (the output $F_w$ can be measured by strain gauges). If $\alpha^{\text{ref}} =\alpha^{\text{trim}}$, the shear force references are identical to their trimmed values, and any variations caused by disturbances or maneuvers (such as the lift drop during a sharp-roll) will be automatically counteracted by flaps. 

Furthermore, the reference generation algorithm for the left and right wing-root bending moments are given in Algorithm~1. 

\begin{algorithm}[H]
\SetAlgoLined
 \begin{algorithmic}
\STATE 
$M_{\phi,r}^{\text{ref}} \gets M_{\phi,r}^{\text{trim}} -  M_{\phi,\text{diff}}^{\text{ref}}/2$; ~~~~~
     $M_{\phi,l}^{\text{ref}} \gets M_{\phi,l}^{\text{trim}} + M_{\phi,\text{diff}}^{\text{ref}}/2$
\IF {$M_{\phi,r}^{\text{ref}} >  \overline {M}_{\phi}^{\text{ref}} $} 
        \STATE $ M_{\phi,r}^{\text{ref}} \gets \overline {M}_{\phi}^{\text{ref}};~~~~~M_{\phi,l}^{\text{ref}} \gets  M_{\phi,\text{diff}}^{\text{ref}} + \overline {M}_{\phi}^{\text{ref}}$
\ELSE
        \IF {$M_{\phi,l}^{\text{ref}} >  \overline {M}_{\phi}^{\text{ref}} $}
              \STATE $ M_{\phi,r}^{\text{ref}} \gets  \overline {M}_{\phi}^{\text{ref}} - M_{\phi,\text{diff}}^{\text{ref}};~~~~~M_{\phi,l}^{\text{ref}} \gets \overline {M}_{\phi}^{\text{ref}}$
        \ENDIF
\ENDIF 
\end{algorithmic}
 \caption{Reference generation for left and right wing-root bending moments}
\end{algorithm}

In Algorithm~1, $M_{\phi,\text{diff}}^{\text{ref}}$ has been designed in subsection~\ref{sub_set_attitude}, which is the core for achieving the designed roll angle tracking performance. $\overline {M}_{\phi}^{\text{ref}}$ in the upper limit on the reference for wing-root bending moment ${M}_{\phi}$, which can be designed \textit{much smaller} than the real \textit{physical} limit $\max( {M}_{\phi})$, i.e., $M_{\phi,l}^{\text{trim}} =M_{\phi,r}^{\text{trim}} < \overline {M}_{\phi}^{\text{ref}} < \max( {M}_{\phi}) $. This design can ensure a safety margin from $\max( {M}_{\phi})$, and is also able to constrain the variations of ${M}_{\phi}$ for extending structural fatigue life.

As formulated in Algorithm~1, to achieve a moderate roll maneuver, the left and right wing take the same responsibility. However, once one of the wing-root bending moments reaches $\overline {M}_{\phi}^{\text{ref}}$ under a sharp-roll command, the exceeded command will be automatically allocated to the other half-wing. In such a way, not only the wing loads are limited, $M_{\phi,\text{diff}}^{\text{ref}}$ is achieved for roll-command tracking as well. 

\begin{remark}
\rm The wing-root shear force and bending moment reference generation algorithms use very little model information, and are easy to be implemented in real-time. More importantly, by exploiting the control redundancy, the load alleviation is achieved without degrading the command tracking performance.
\end{remark}

After the references $F_{w,r}^{\text{ref}},F_{w,l}^{\text{ref}} , M_{\phi,r}^{\text{ref}},M_{\phi,l}^{\text{ref}}$ are derived, a controller should be designed to track these references. Taking the right wing as an example, only two independent control variables are needed for tracking $F_{w,r}^{\text{ref}}$ and $M_{\phi,r}^{\text{ref}}$. Since we have seven flaps on each wing, the remaining control space can be used for flutter suppression, aeroelastic damping enhancement, and control energy reduction. Choose the controlled output as $\boldsymbol{y} =[F_{w,r}, M_{\phi,r}]^\mathsf{T}$, then the right-wing dynamics given by Eqs.~\eqref{eq_wing_aero} and~\eqref{eq_F_root} are represented as
\begin{eqnarray}
    \dot {\boldsymbol{x}}_e &=& \boldsymbol{A} {\boldsymbol{x}}_e +\boldsymbol{B}_u \boldsymbol{u} + \boldsymbol{B}_g\boldsymbol{\alpha}_g   + \boldsymbol{B}_{\text{grav}} + \boldsymbol{B}_{r} \boldsymbol{\alpha}_{qs,r} + \boldsymbol{B}_{\text{acc}} \nonumber \\
    \boldsymbol{y} &=& \boldsymbol{C} {\boldsymbol{x}}_e +\boldsymbol{D}_u \boldsymbol{u}  + \boldsymbol{D}_{\text{grav}} + \boldsymbol{D}_{r} \boldsymbol{\alpha}_{qs,r} + \boldsymbol{D}_{\text{acc}}
\end{eqnarray}

Further design an augmented state vector $\boldsymbol{X}= [ {\boldsymbol{x}}_e^\mathsf{T}, \int \boldsymbol e_{y}^\mathsf{T} ]^\mathsf{T}$, $\boldsymbol e_{y} = \boldsymbol{y} - \boldsymbol{y}^{\text{ref}}$ for load commands tracking, then the augmented system dynamics are
\begin{equation}
  \begin{bmatrix}
  \dot {\boldsymbol{x}}_e \\ 
  \boldsymbol e_{y} 
  \end{bmatrix}  
  = \begin{bmatrix}
  \boldsymbol{A}  & \boldsymbol{0} \\
    \boldsymbol{C} & \boldsymbol{0}
  \end{bmatrix}
  \begin{bmatrix}
  {\boldsymbol{x}}_e \\  
  \int  \boldsymbol e_{y}  
  \end{bmatrix}+
  \begin{bmatrix}
  \boldsymbol{B}_u \\
  \boldsymbol{D}_u
  \end{bmatrix} \boldsymbol{u}  +
  \begin{bmatrix}
  \boldsymbol{B}_r \\
  \boldsymbol{D}_r
  \end{bmatrix} \boldsymbol{\alpha}_{qs,r} +
  \begin{bmatrix}
  \boldsymbol{B}_{\text{g}} \\
  \boldsymbol{0}
  \end{bmatrix} \boldsymbol{\alpha}_{g}+  \begin{bmatrix}
  \boldsymbol{B}_{\text{grav}} \\
  \boldsymbol{D}_{\text{grav}}
  \end{bmatrix} + 
  \begin{bmatrix}
  \boldsymbol{B}_{\text{acc}} \\
  \boldsymbol{D}_{\text{acc}}
  \end{bmatrix} + 
  \begin{bmatrix}
  \boldsymbol{0} \\
  - \boldsymbol {y}^{\text{ref}}
  \end{bmatrix}
\end{equation}
    
To evaluate the robustness of the controller, the disturbance input $\boldsymbol{\alpha}_g$ and the inertial input vector $[\boldsymbol{B}_{\text{acc}}^\mathsf{T},  \boldsymbol{D}_{\text{acc}}^\mathsf{T}]^\mathsf{T}$ are eliminated from the model used for control design. The control performance can be further enhanced if extra knowledge about $\boldsymbol{\alpha}_g$ is available (from LIDAR or a gust estimator). The local quasi-steady angle of attack induced by rigid-body translational and rotational motion $\boldsymbol{\alpha}_{qs,r}$ (Eq.~\eqref{input_alpha}) is treated as a known input in wing control design. In addition, the gravitational forces are also treated as known inputs in wing control design. Formulate a cost function as 
\begin{equation}
    J = \lim_{t_f \rightarrow \infty } \frac{1}{2}\int_0^{t_f} [ \boldsymbol{X}^\mathsf{T} \boldsymbol{ Q} \boldsymbol{X} + \boldsymbol{u}^\mathsf{T} \boldsymbol{R} \boldsymbol{u}]~ \text{d}t
\end{equation}
where $\boldsymbol{ Q}$ and $\boldsymbol{ R}$ are positive definite diagonal matrices. Denote the augmented system dynamic matrix as $\boldsymbol A_{\text{aug}}$, and denote $\boldsymbol B_{\text{aug}} = [\boldsymbol{B}_u^\mathsf{T},\boldsymbol{D}_u^\mathsf{T}]^\mathsf{T}$, then the infinite time-horizon optimal control is designed as 
\begin{equation}
    \boldsymbol{u} = \boldsymbol{K}_X \boldsymbol{X} + \boldsymbol{K}_r \left(   \begin{bmatrix}
  \boldsymbol{B}_r \\
  \boldsymbol{D}_r
  \end{bmatrix} \boldsymbol{\alpha}_{qs,r} +   \begin{bmatrix}
  \boldsymbol{B}_{\text{grav}} \\
  \boldsymbol{D}_{\text{grav}}
  \end{bmatrix} + 
  \begin{bmatrix}
  \boldsymbol{0} \\
  - \boldsymbol {y}^{\text{ref}}
  \end{bmatrix}\right)
\end{equation}
where 
\begin{equation}
    \boldsymbol{K}_X = - \boldsymbol{R}^{-1} \boldsymbol{B}_{\text{aug}}^\mathsf{T}\boldsymbol{W},~~~ \boldsymbol{K}_r = - \boldsymbol{R}^{-1} \boldsymbol{B}_{\text{aug}}^\mathsf{T}(\boldsymbol{W}\boldsymbol{B}_{\text{aug}} \boldsymbol{R}^{-1} \boldsymbol{B}_{\text{aug}} ^\mathsf{T}- \boldsymbol{A}_{\text{aug}}^\mathsf{T} )^{-1}\boldsymbol{W}
    \label{lqr_u}
\end{equation}

\begin{figure}[!h]
	\begin{center}
		\includegraphics[width=0.99\linewidth]{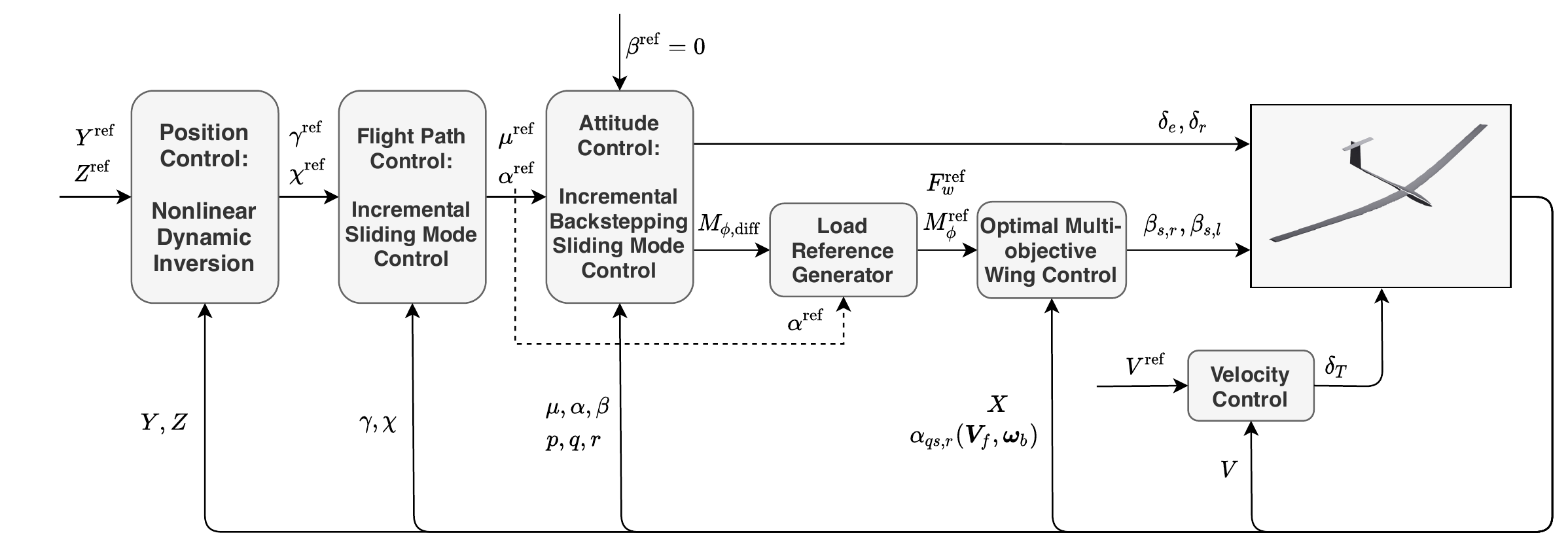}
		\caption{Control architecture.}
  		\label{block_diagram}
	\end{center}
\end{figure}

$\boldsymbol{W}$ in Eq.~\eqref{lqr_u} is the solution of the associated Riccati equation. The designed control input can achieve the following goals: 1) stabilize the wing aeroelastic system (flutter suppression); 2) enhance structural damping (by assigning the closed-loop eigenvalues); 3) track the designed load commands; 4) minimize control energy. The states needed for feedback control can be observed by a Kalman filter~\cite{Wang2019c}. The proposed control architecture is illustrated in Fig.~\ref{block_diagram}.

\section{Simulation Results and Discussions}
\label{sec_results}

\begin{figure}[!h]
	\begin{center}
		\includegraphics[width=.45\linewidth]{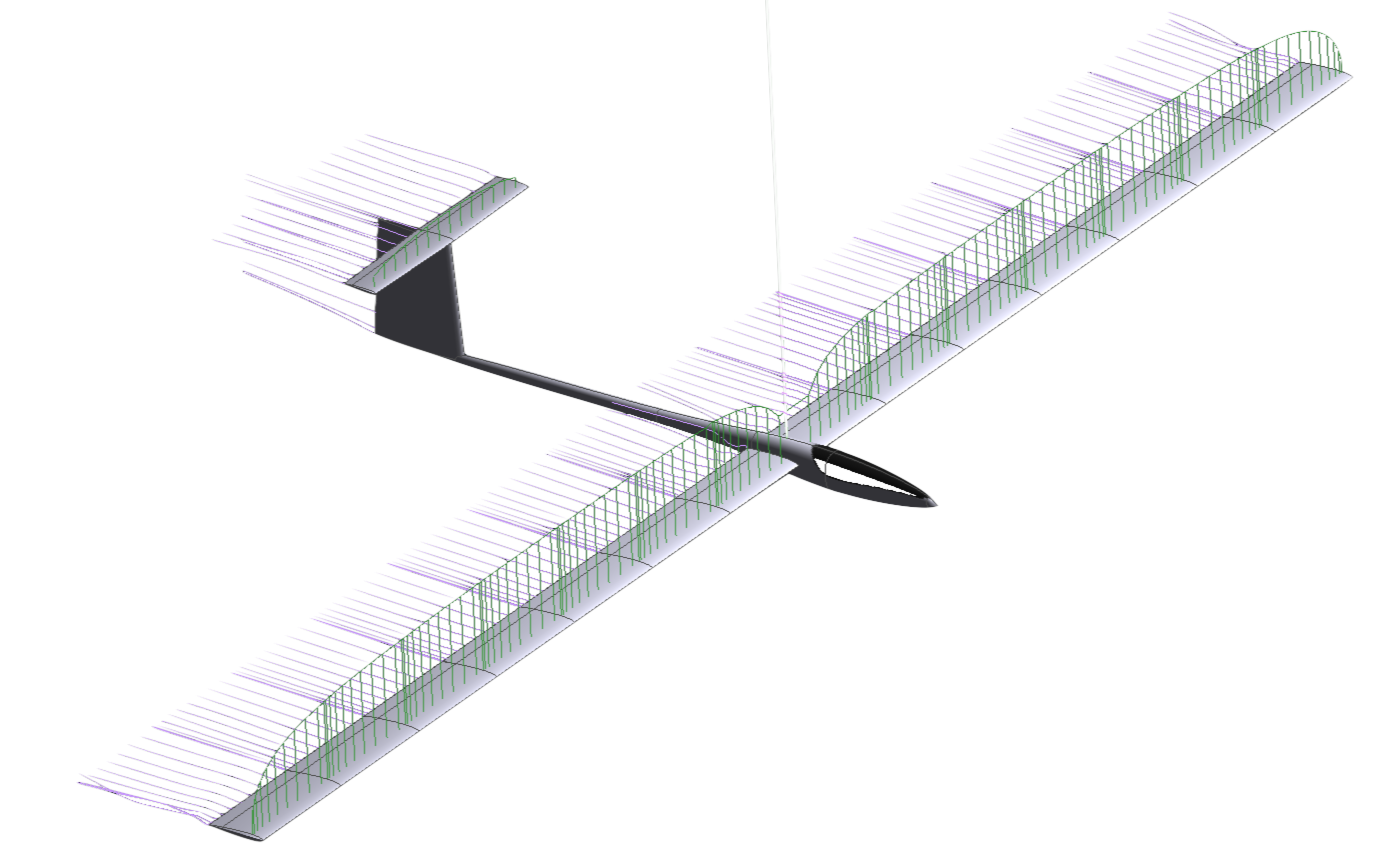}
		\caption{3D view of the flexible aircraft model.}
  		\label{fig:plane3d}
	\end{center}
\end{figure}

\begin{figure}[!h]
\centering
\begin{subfigure}[t]{0.33\textwidth}
	\includegraphics[width=\textwidth]{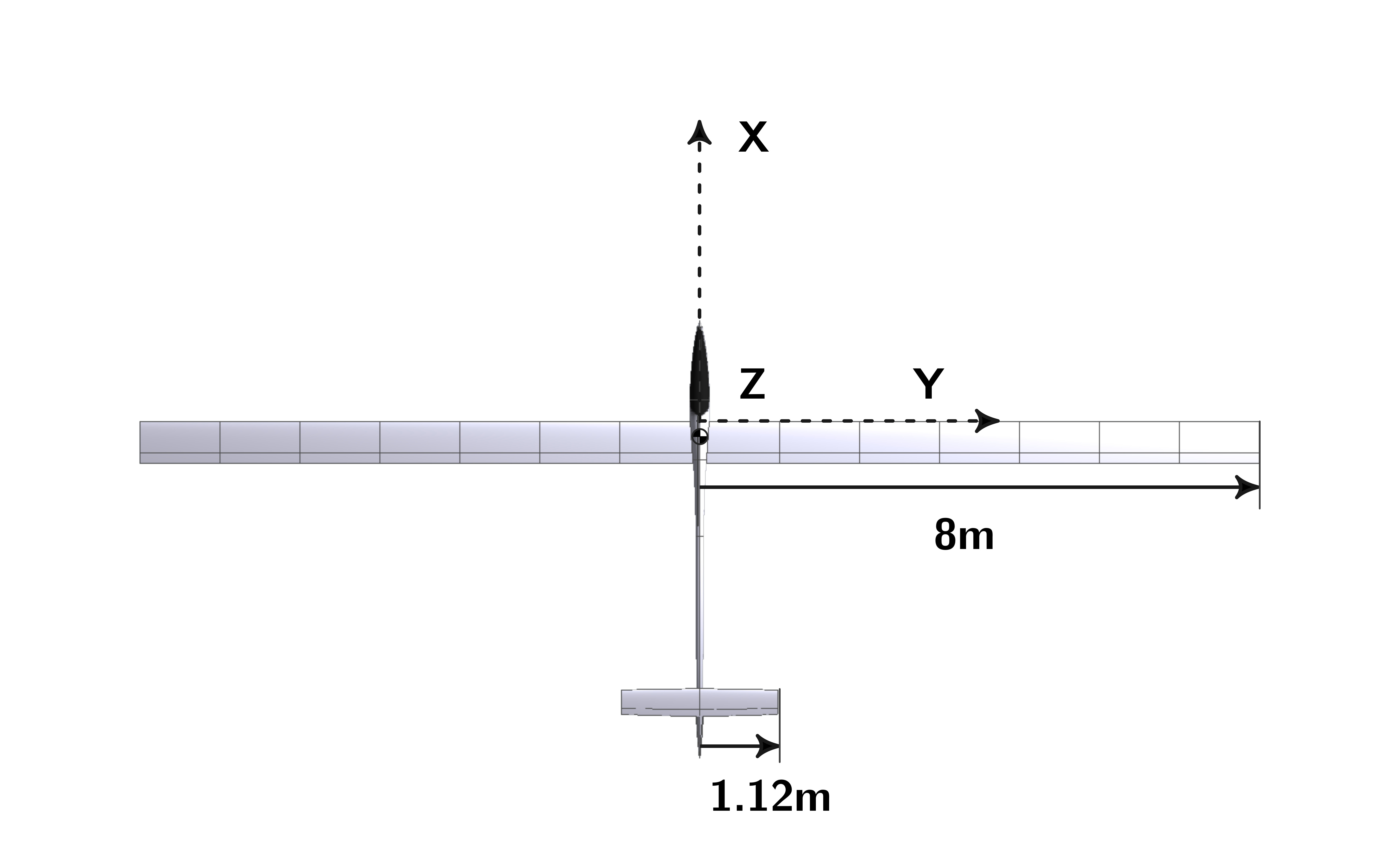}
	\caption{Top view $(x,y)$}
	\label{fig:plane_top}
\end{subfigure}
\hfill
  \begin{subfigure}[t]{0.33\textwidth}
    \includegraphics[width=\textwidth]{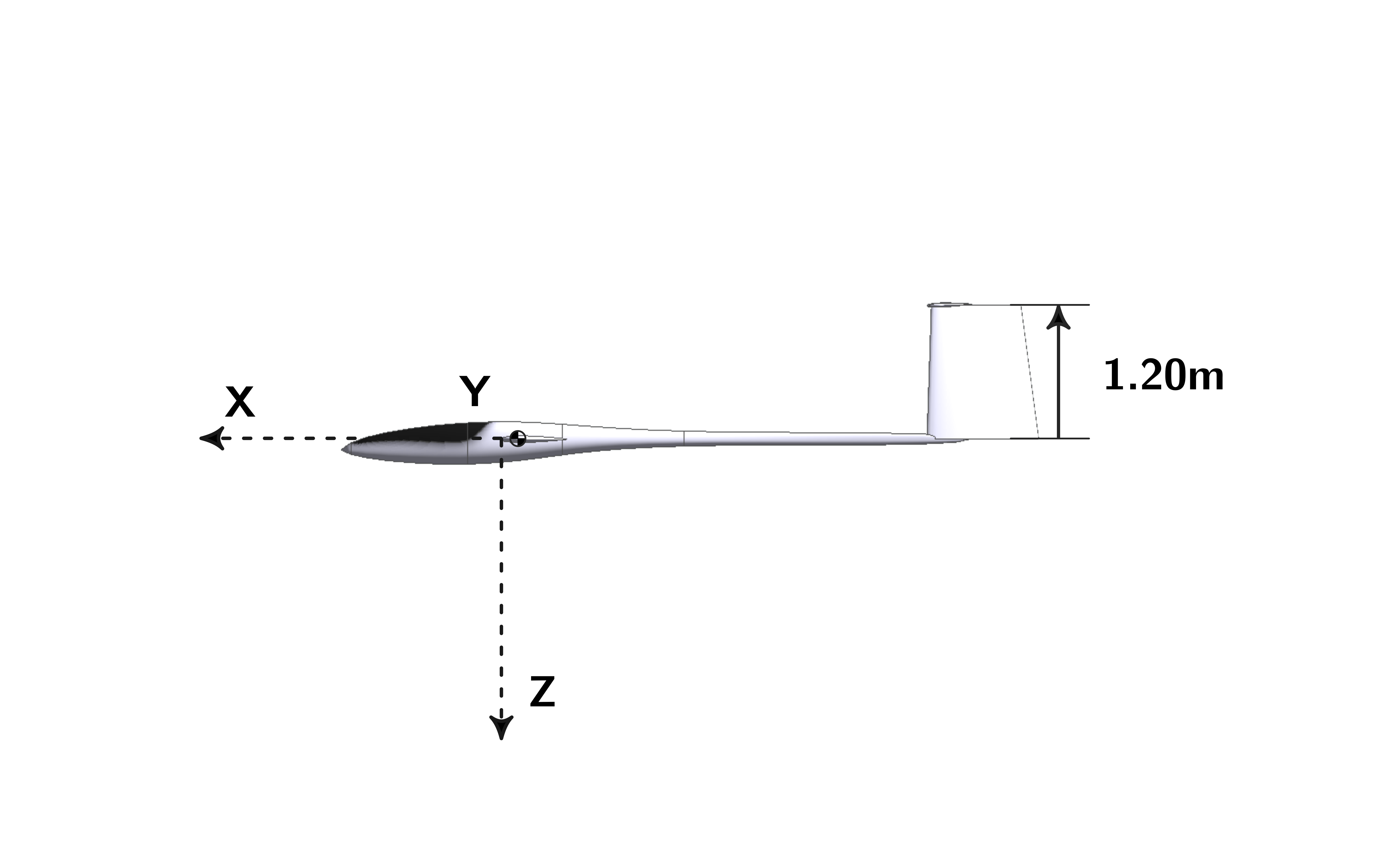}
    \caption{Side view $(x,z)$}
    \label{fig:plane_back}
  \end{subfigure}
  \hfill
    \begin{subfigure}[t]{0.33\textwidth}
    \includegraphics[width=\textwidth]{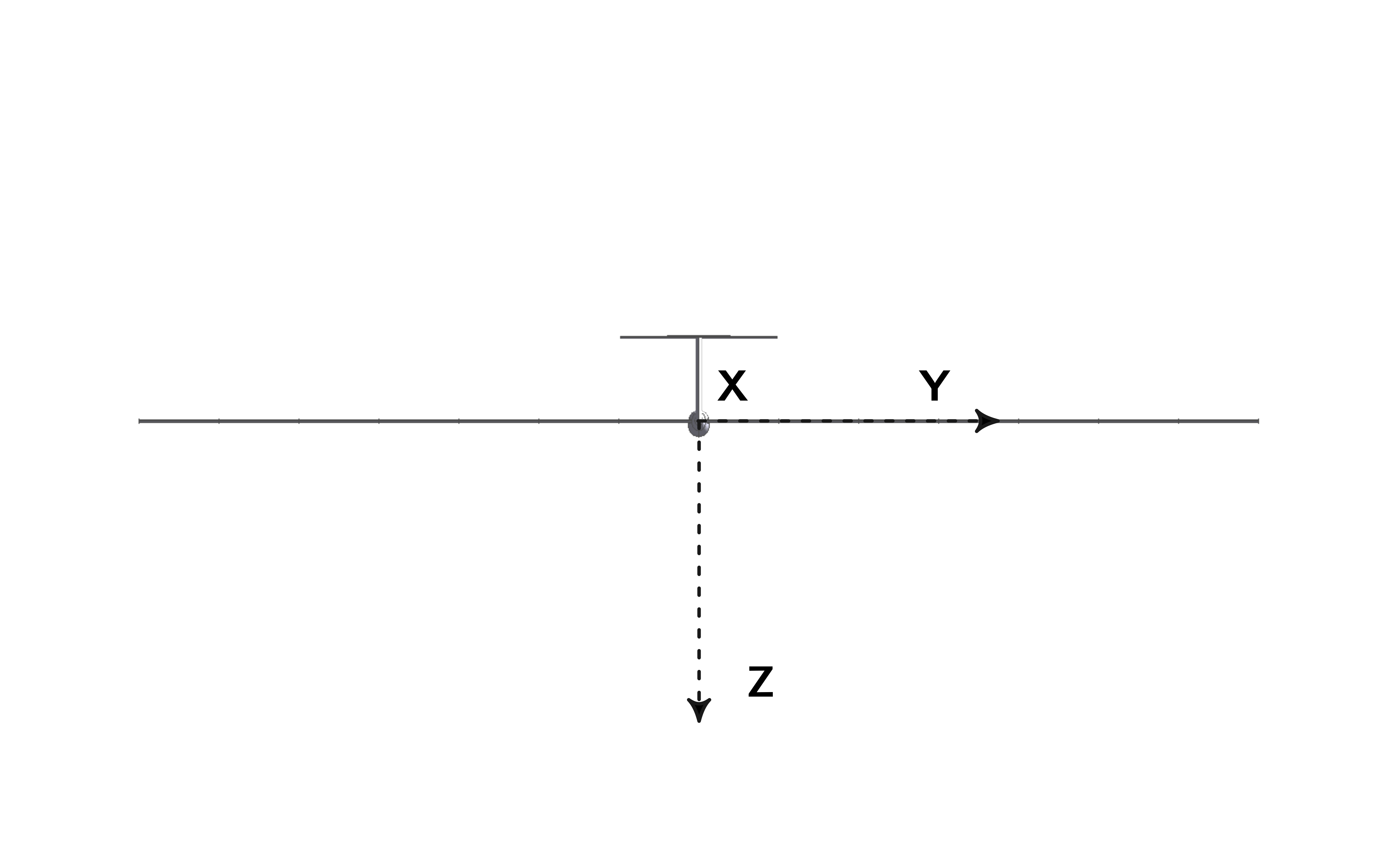}
    \caption{Back view $(y,z)$}
    \label{fig:plane_top}
  \end{subfigure}
  \caption{Top, back and side views of the flexible aircraft.}
  \label{fig:planform}
\end{figure}

In this section, the effectiveness of the proposed control architecture is evaluated. A 3D view of the planform is shown in Fig.~\ref{fig:plane3d}. Top, back and side views of the planform are shown in Fig.~\ref{fig:planform}. The wing was designed using XFLR5. The steady-state aerodynamic lift coefficients were extracted using the vortex lattice method with a total of 1967 mesh elements. The aspect ratio of the wing equals 26.67. The total mass of the aircraft equals 227~kg. The aircraft moment of inertia in trim condition is $I_{xx} =493.8$~kg$\cdot$m$^2$, $I_{yy} =726.7$~kg$\cdot$m$^2$, $I_{zz} =1170.5$~kg$\cdot$m$^2$, and $I_{xz} =-34.0$~kg$\cdot$m$^2$. During maneuvers, these inertia will change with structural deformations. The actuator dynamics of elevator and rudder are modeled as first-order systems, with time constant equals 0.02~s. The deflection limits of them both equal $\pm 20$~deg. As discussed in subsection~\ref{subsection_ase_model}, the fourteen wing flaps are connected with the main wing structure via rotational springs. The inertia couplings between the flap and structure are considered in the structure mass matrix. The deflection limits of the flaps are set as $\pm 30$~deg. In addition, the throttle dynamics are model as a first-order system with time constant equals 0.2~s, which is slower than that of control surfaces. To capture the high-frequency dynamics of the wing aeroelastic dynamics, the simulation sampling frequency is chosen as 20,000~Hz. On the contrary, the sampling frequency used by controllers are much lower. The sampling frequency used by the attitude and wing control loops is 100~Hz, while 50~Hz is used by the position and flight path control loops. 100~Hz is a reasonable choice for the incremental control of aircraft attitude dynamics, which has been verified by flight tests on a CS-25 aircraft~\cite{Grondman2018}.

\subsection{Trim and Model Analysis}

A free-flying rigid aircraft model is set up for comparative studies. This rigid aircraft shares the same geometric and static aerodynamic properties with the free-flying flexible aircraft model. Nevertheless, its structural stiffness are assumed to be infinitely high, and its aerodynamic model relies on quasi-steady strip theory. Both the rigid and flexible aircraft are trimmed in a steady-level flight condition at $H = 1000$~m, $V = 35$~m/s (a typical operational point for glider aircraft). The trim solutions are shown in Table~\ref{trim_tabb}.  
\begin{table}[!h]
 \begin{center}
  \caption{Trim solutions for the rigid and flexible aircraft.}
  \label{trim_tabb}
  \begin{tabular}{lccc}
  \hline\hline
        & $\alpha_*~[^\circ]$ & $\delta_{e_*}~[^\circ]$ &  $F_{E_*}$ [N]  \\ \hline
        Rigid aircraft &  3.467& $-0.1578$ & 109.9 \\ 
        Flexible aircraft &3.626 & $-0.1337$ & 101.5 \\
       \hline
        \hline
  \end{tabular}
 \end{center}
\end{table}

\begin{figure}[!h]
  \centering
  \includegraphics[width=1\textwidth]{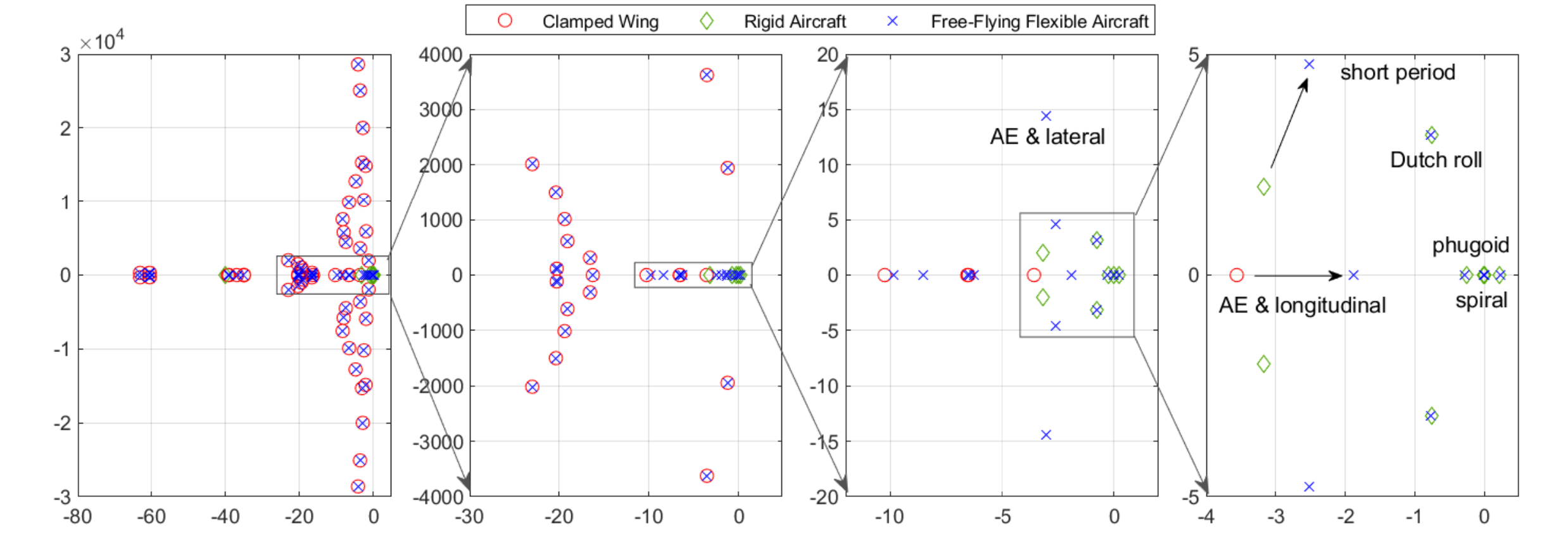}
  \caption{Eigenvalues of the aereoelastic clamped wing, the rigid aircraft, and the free-flying flexible aircraft.}
  \label{eig_compare}
\end{figure}

To analyze their characteristics, the rigid and flexible aircraft are linearized around their equilibrium points, with the resulting eigenvalues illustrated in Fig.~\ref{eig_compare}. The poles of the clamped aeroelastic wing are also shown. It can be seen from Fig.~\ref{eig_compare}, most of the flexible aircraft poles appear in the high-frequency range, and are in good agreement with those of the clamped wing. Discrepancies caused by the interactions between structural and rigid-body motions are more visible in the low-frequency range, which can be seen in the third sub-polt of Fig.~\ref{eig_compare}. A new periodic mode with natural frequency 14.38 rad/s, and damping ratio 0.21 also emerges. The corresponding eigenvectors indicate this mode is dominated by the couplings between wing bending and rigid-body roll rate. The fourth sub-plot shows the frequency range of the conventional rigid-aircraft. However, an aperiodic mode of the aeroelastic clamped wing appears in this frequency range. Because of the interactions, the flexible aircraft short-period damping ratio is only 0.46, which is almost halved as compared to its rigid counterpart (0.84). Moreover, the couplings also make the aperiodic mode move towards the unstable region. An eigenvector analysis shows the pole at -1.876 is dominated by coupled aircraft longitudinal motions and wing aeroelastic modes. In addition, the Dutch roll, phugoid and spiral modes of the flexible aircraft are in the vicinity of their rigid counterparts.

The rigid-body and structural couplings are clearly exposed in the above analyses, which underline the necessity of a multi-objective integrated controller. From Fig.~\ref{eig_compare}, the phugoid mode should be stabilized. Moreover, the damping ratio of the short-period, Dutch roll and the new aeroelastic-lateral coupled modes need to be increased. Furthermore, the controller should simultaneously fulfill the trajectory tracking commands and the load alleviation requirements. The effectiveness of the proposed control architecture will be shown in the following subsections.

\subsection{Maneuver Load Alleviation}

Four maneuver circumstances will be considered in this subsection: sudden pull-up, sharp roll, flight path control, and 3D aircraft position control. The controller aims at minimizing the tracking errors while alleviating the loads caused by maneuvers.

\subsubsection{Load Alleviation in a Pull-up Maneuver}

\begin{figure}[!h]
  \centering
  \includegraphics[width=0.55\textwidth]{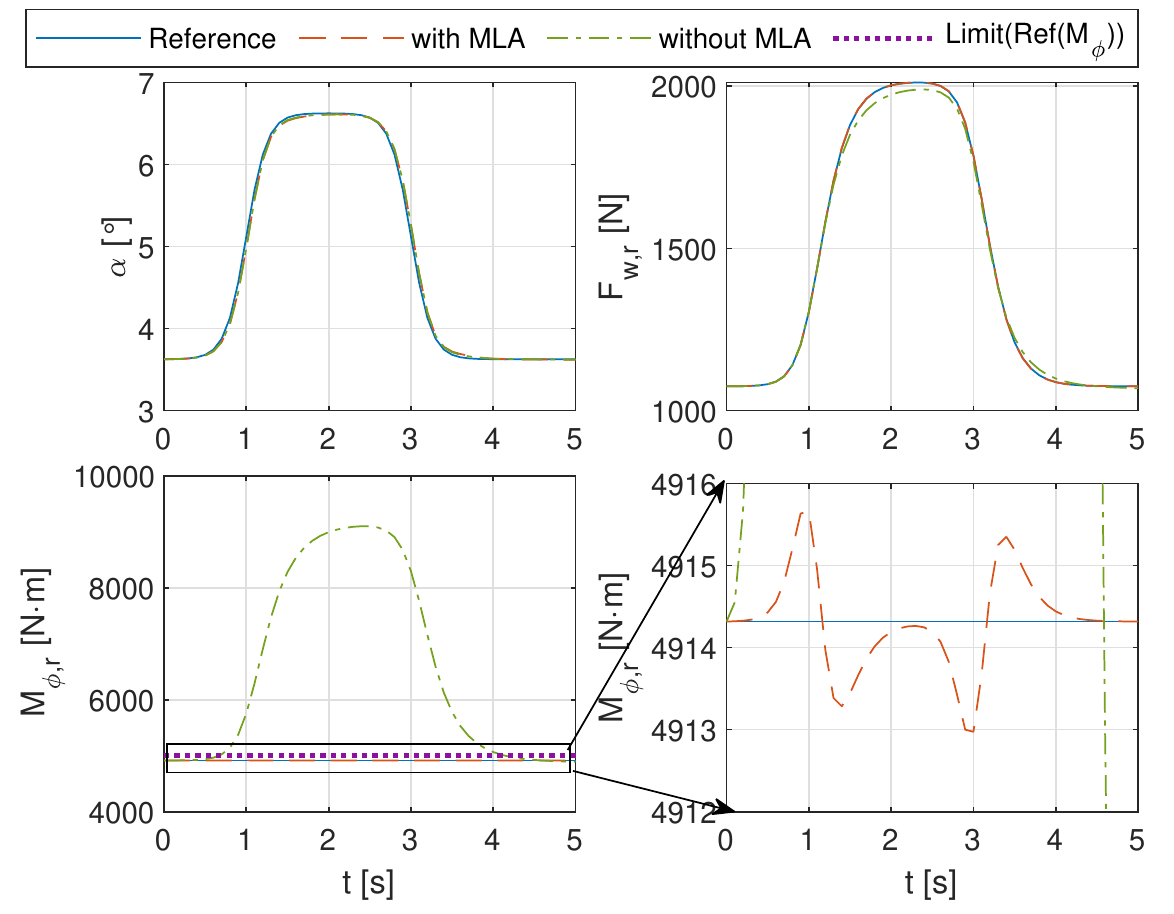}
  \caption{$\alpha$ tracking performance and the responses of wing-root shear force and bending moment.}
  \label{alpha_loads}
\end{figure}

In this simulation case, the aircraft is commanded to track an $\alpha$ profile, the kinematic bank angle and the side-slip angle are commanded to remain zero. As illustrated in the first subplot of Fig.~\ref{alpha_loads}, two smoothly combined sigmoid functions $f_1 = {1}/(1+e^{-8(t-1)})$ and $f_2 = -{1}/(1+e^{-8(t-3)})$ are used to compose the command. In practice, a common choice of command is a discontinuous step function filtered by a low-pass filter. However, at the step points, the filtered signal is still non-differentiable. These non-differentiable points would cause spikes in control inputs. In view of this, the sigmoid function, which is differentiable up to any order, is chosen as a smooth realization of the step function in this paper. Since this maneuver is symmetrical, only the right wing responses are shown. 

\begin{figure}[!h]
  \centering
  \includegraphics[width=0.55\textwidth]{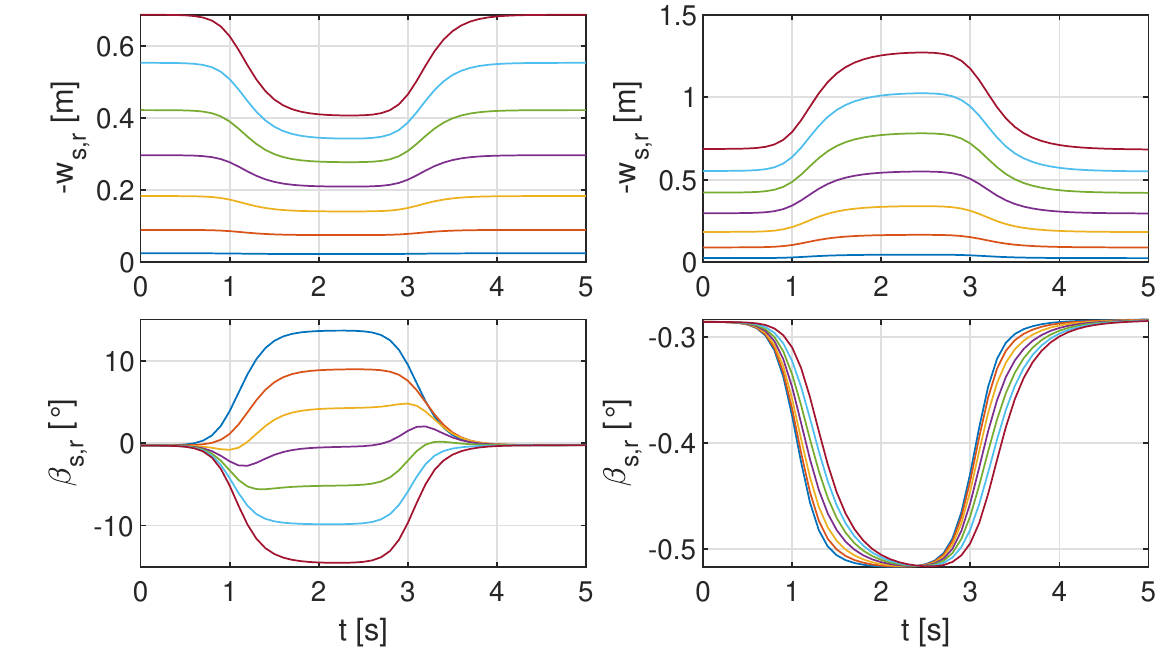}
  \caption{Responses of the wing vertical displacements and flap angles with (left) and without MLA (right).}
  \label{alpha_wing}
\end{figure}

To demonstrate the effectiveness of the proposed control law, another controller without the MLA function is also designed. These two controllers give the same commands to elevator and rudder. However, for the controller without MLA, the flap hinge moment commands are set to zero during symmetrical maneuvers. As shown in Fig.~\ref{alpha_loads}, both the controllers with and without MLA can track the given $\alpha$ command, with $\max(|\alpha - \alpha^{\text{ref}}|)<0.14^\circ$. The wing-root shear force responses of the two controllers are also similar. Nevertheless, the wing-root bending moment is effectively reduced by the MLA function. Recall Algorithm~1, $M_{\phi,r}^{\text{ref}} = M_{\phi,r}^{\text{trim}}$ in symmetrical maneuvers ($M_{\phi,\text{diff}}^{\text{ref}}=0$). As shown in the third and fourth subplots, the usage of MLA reduces $\max(|M_{\phi,r} - M_{\phi,r}^{\text{trim}}|)$ from 4183~N$\cdot$m to 1.344~N$\cdot$m (by 99.97\%), and reduces $\text{rms}(|M_{\phi,r} - M_{\phi,r}^{\text{trim}}|)$ from 2387~N$\cdot$m to 0.5827~N$\cdot$m (by 99.98\%). It can also be seen from Fig.~\ref{alpha_wing}, that the MLA function make the inner-board flaps deflect downwards, and the out-board flaps deflect upwards, which shifts the pressure center towards the wing root. More importantly, by exploiting the control redundancy, the $\alpha$ tracking performance is not influenced by the MLA function. 

\subsubsection{Load Alleviation in a Sharp-roll Maneuver}

In this test case, the aircraft is commanded to roll from 0~deg to 40~deg within two seconds. The sigmoid  function $f_3 = {1}/(1+e^{-6(t-2)})$ is adopted as a smooth realization of the step function. Another controller without the MLA function is designed for comparisons. This controller uses the same outer-loop control algorithms with the nominal controller, but there are two differences in the optimal wing control loop: 1) the shear force command tracking is eliminated; 2) the right and left wing always share the same responsibility in roll tracking, i.e., $M_{\phi,r}^{\text{ref}} \equiv M_{\phi,r}^{\text{trim}} -  M_{\phi,\text{diff}}^{\text{ref}}/2,~ M_{\phi,l}^{\text{ref}} \equiv M_{\phi,l}^{\text{trim}} + M_{\phi,\text{diff}}^{\text{ref}}/2$. Apart from these two differences, the controller without MLA still uses the same cost function and gains with the nominal controller. 

\begin{figure}[!h]
  \centering
  \includegraphics[width=0.55\textwidth]{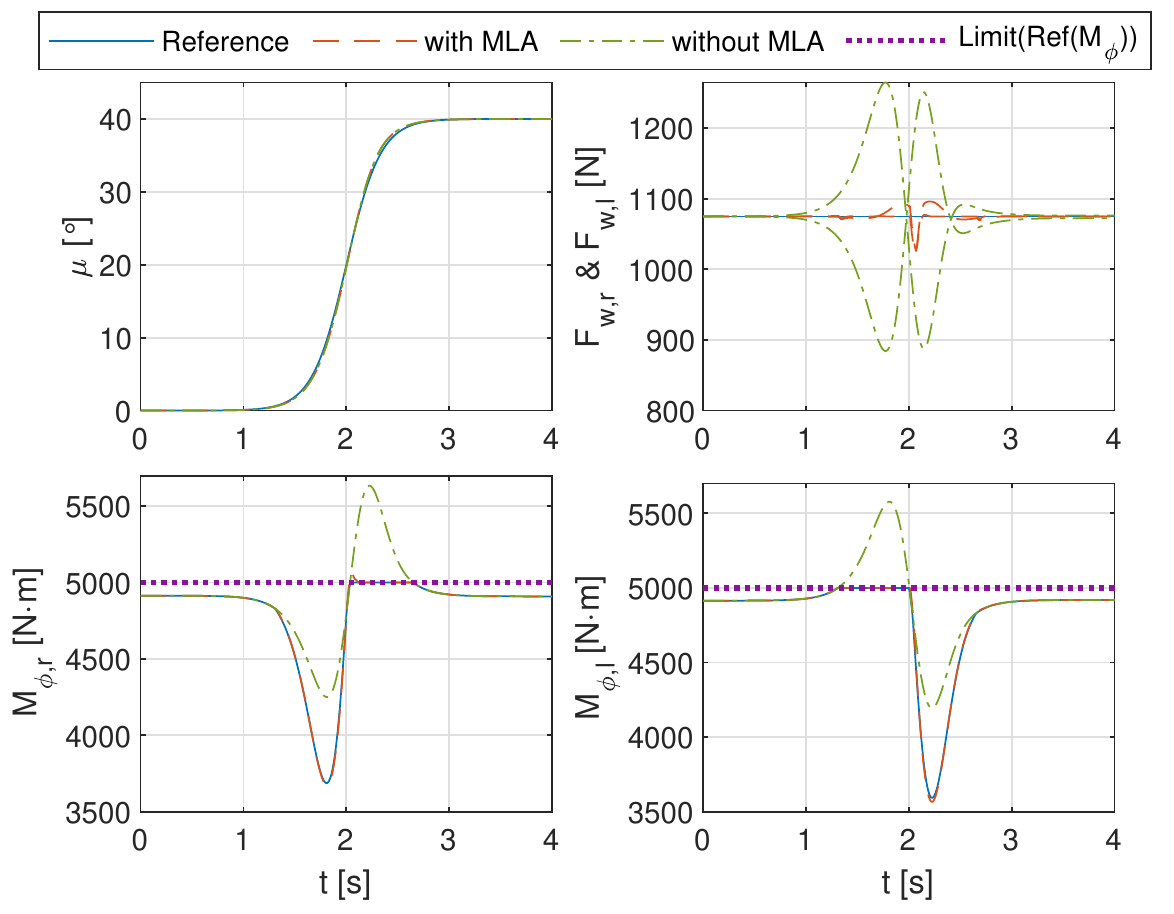}
  \caption{Responses of the kinematic bank angle, wing-root shear force and bending moment.}
  \label{sharp_roll_loads}
\end{figure}

In order to demonstrate the effectiveness of the proposed control architecture, the upper limit on the reference for wing-root bending moment $\overline {M}_{\phi}^{\text{ref}}$ is set at 5000~N$\cdot$m, which is \textit{much lower} than the real physical limit on ${M}_{\phi}$. In the steady-level flight, the trimmed value for root bending moment equals 4914~N$\cdot$m, thus $|\overline {M}_{\phi}^{\text{ref}}- M_{\phi}^{\text{trim}}|$ is less than 2\% of $M_{\phi}^{\text{trim}}$. Under this strict constrain, and tracking performance and load responses are shown in Fig.~\ref{sharp_roll_loads}.

In view of Fig.~\ref{sharp_roll_loads}, the kinematic bank angle tracking performance with and without MLA are almost identical. Both of them can track the given command very well, with $\max(|\mu - \mu^{\text{ref}}|)<0.66^\circ$. The responses of the left and right wing-root shear forces are shown in the second subplot, from which it can be seen that without MLA, $F_{w,r}$ and $F_{w,l}$ have large variations because of the $\boldsymbol{\alpha}_{qs,r}$ induced by roll maneuvers. By contrast, the MLA function can compensate for the shear force variations, since $F_{w,r}^{\text{ref}} = F_{w,l}^{\text{ref}} = F_{w}^{\text{trim}}$ when $\alpha^{\text{ref}} = 0$ (Eq.~\eqref{F_w_ref}). When using the MLA function, $\text{rms}(F_{w,r}-F_{w,r}^{\text{ref}})$ is reduced from 63.83~N to 6.235~N (by 90.23\%). Moreover, as illustrated in the third and fourth subplots, the MLA algorithm successfully complies with the strict limit $\overline {M}_{\phi}^{\text{ref}} =5000$~N$\cdot$m. Once one of ${M}_{\phi,r},~{M}_{\phi,l}$ reaches $\overline {M}_{\phi}^{\text{ref}}$, the exceeded command is automatically allocated to the other wing without influencing the roll tracking performance. On the contrary, if the MLA algorithm is not used during this maneuver, $\text{max}({M}_{\phi,r})= 5634$~N$\cdot$m, and $\text{max}({M}_{\phi,l})=  5578$~N$\cdot$m.

\begin{figure}[!h]
  \centering
  \includegraphics[width=0.55\textwidth]{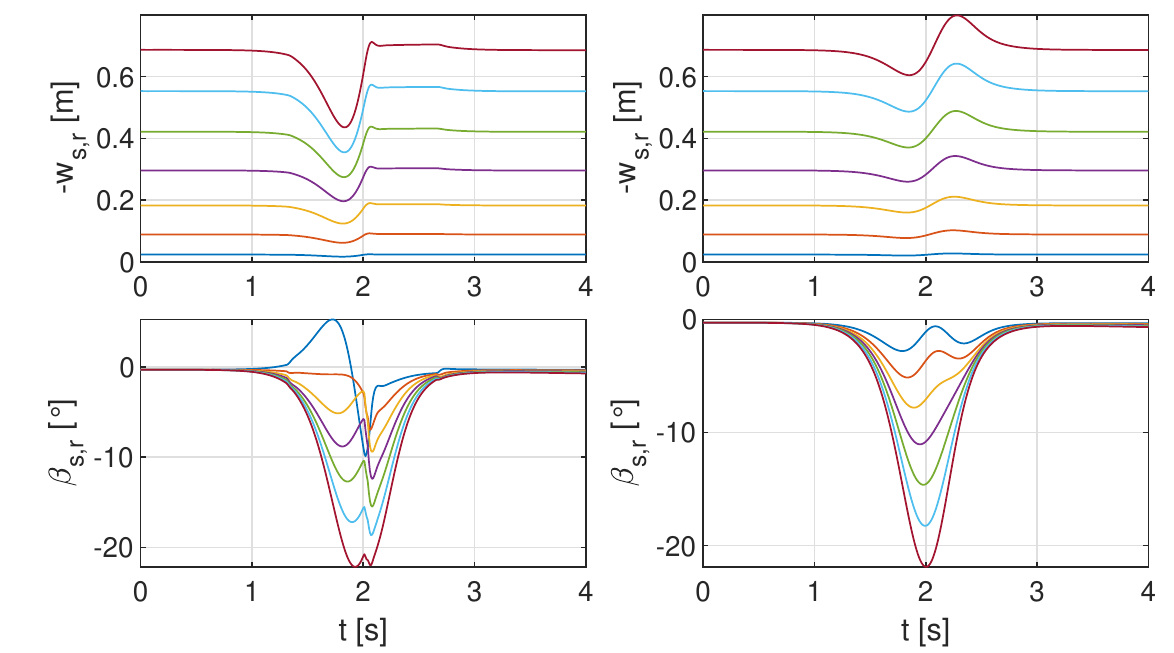}
  \caption{Responses of the \textit{right} wing displacements and flap deflections with (left) and without MLA (right).}
  \label{sharp_roll_wing_right}
\end{figure}

\begin{figure}[!h]
  \centering
  \includegraphics[width=0.55\textwidth]{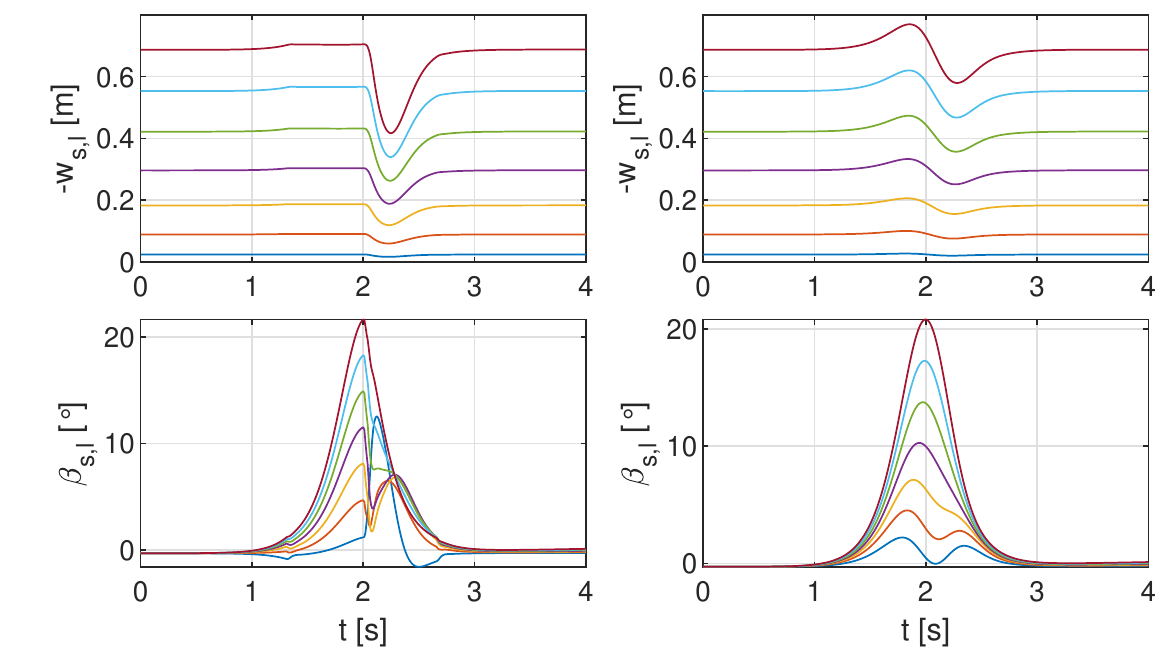}
  \caption{Responses of the \textit{left} wing displacements and flap deflections with (left) and without MLA (right).}
  \label{sharp_roll_wing_left}
\end{figure}

The load reduction without influencing the command tracking performance is achieved by fully exploiting the input redundancy in control design (subsection~\ref{sub_sec_optimal}). The responses of the wing bending displacements and flap deflections during the sharp-roll maneuver are shown in Figs.~\ref{sharp_roll_wing_right} and~\ref{sharp_roll_wing_left}. It can be observed that by actuating the distributed trailing-edge control surfaces, the proposed controller can not only reduce loads, but constrain the maximum wing deflections as well. In addition, the flap deflection angles remain in the $\pm 30$~deg limits.
\subsubsection{Flight Path Control With Maneuver Load Alleviation}

\begin{figure}[!h]
  \centering
  \includegraphics[width=0.5\textwidth]{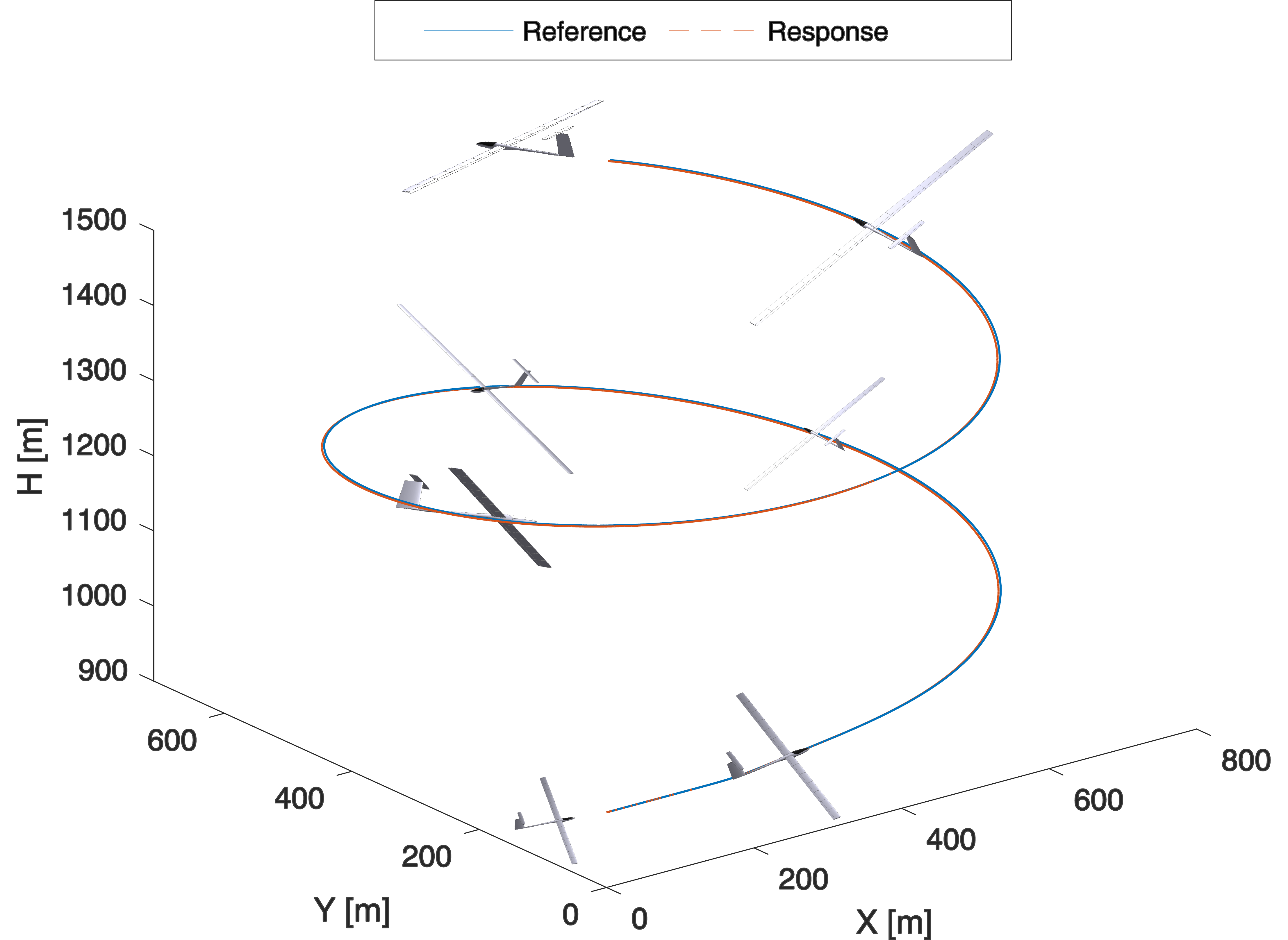}
  \caption{3D spiral trajectory tracking performance.}
  \label{track_sprial_3D}
\end{figure}

\begin{figure}[!h]
  \centering
  \includegraphics[width=0.65\textwidth]{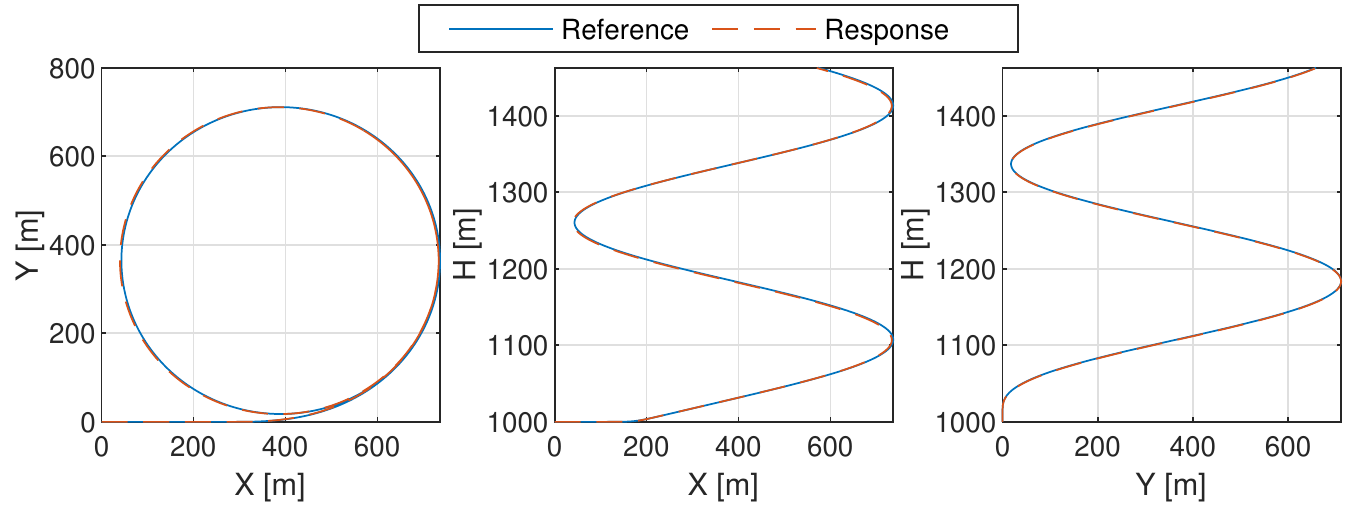}
  \caption{Three views of the 3D spiral trajectory tracking performance.}
  \label{track_sprial_XYZ}
\end{figure}

\begin{figure}[!h]
  \centering
  \includegraphics[width=0.9\textwidth]{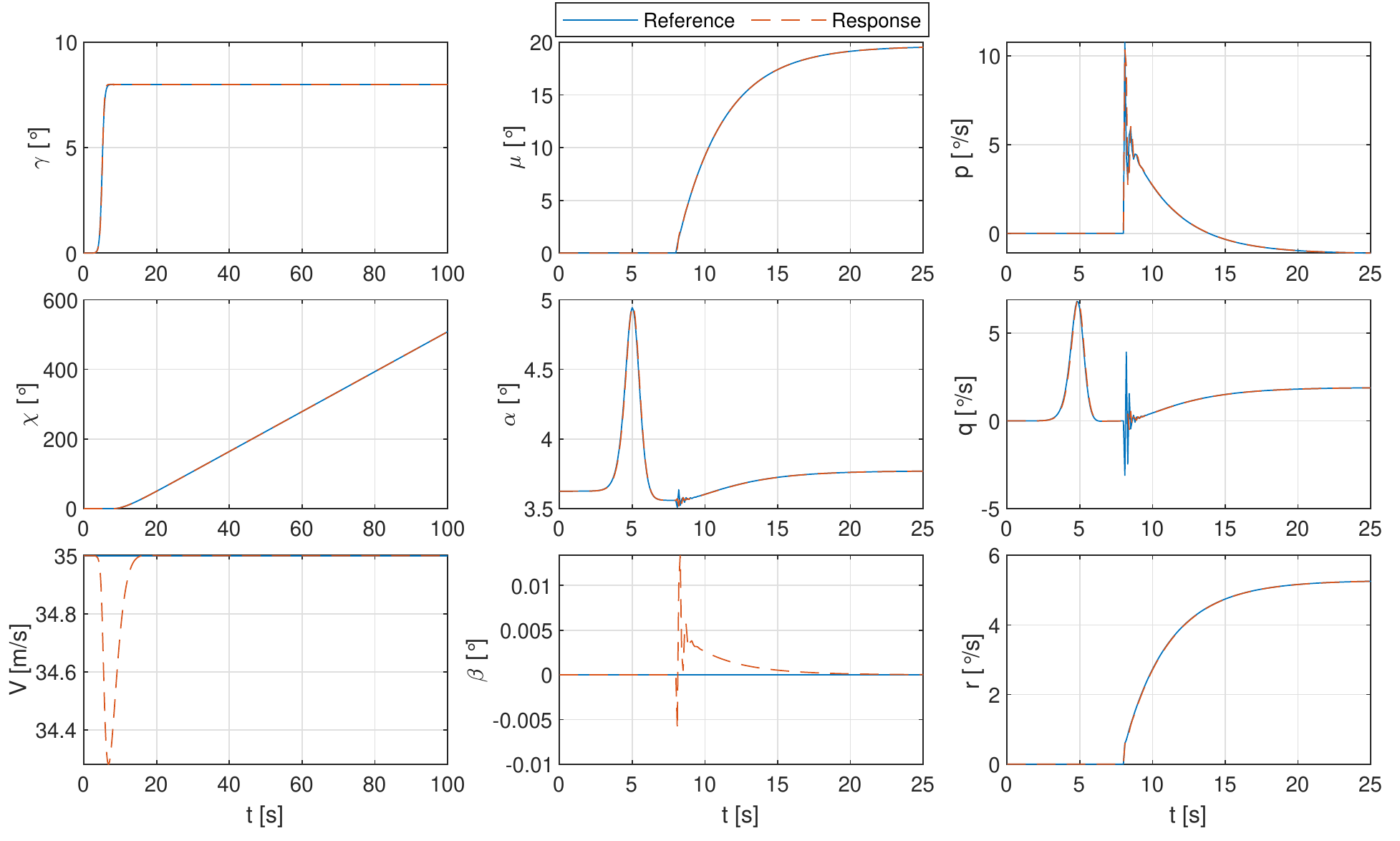}
  \caption{Responses of flight path angles, attitude angles and angular rates in an aircraft spiral maneuver.}
  \label{track_sprial_states}
\end{figure}

In this simulation case, the aircraft is commanded to follow a spiral trajectory, which is shown in Figs.~\ref{track_sprial_3D} and~\ref{track_sprial_XYZ}. This trajectory can be realized by giving a step command to flight path angle $\gamma$, and a ramp command to the kinematic azimuth angle $\chi$. Smooth realizations of the step commands are given to the aircraft, as shown in Fig.~\ref{track_sprial_states}. The upper limit on the reference for wing-root bending moment is still set as $\overline {M}_{\phi}^{\text{ref}} =5000$~N$\cdot$m. 
 
As can be viewed from Figs.~\ref{track_sprial_3D} and~\ref{track_sprial_XYZ}, the proposed controller can make the flexible aircraft track the given command, with $\text{max}(|X - X^{\text{ref}}|) < 3.10$~m, $\text{max}(|Y - Y^{\text{ref}}|) < 0.11$~m and $\text{max}(|H - H^{\text{ref}}|) < 0.42$~m. Fig.~\ref{track_sprial_states} shows the rigid-body state responses during this maneuver. The flight path angle increases from 0~deg to 8~deg within 3.5~s, and the maximum tracking error equals 0.19$^\circ$. A ramp command with slope $5.73^\circ$/s is given to the kinematic azimuth angle $\chi$, and the aircraft can track it with maximum tracking error equals 0.026$^\circ$. The aircraft velocity is maintained at the trimmed value using a linear throttle control ($\delta_T$). Since the aircraft enters a stable spiral mode after $t=25$~s, the responses of inner-loop states are shown in $t\in[0,25]$~s. The attitude commands given by the flight path control loop are also tracked, with $\max(|e_\mu|) = 0.34^\circ$, $\max(|e_\alpha|) = 0.12^\circ$, and $\max(|e_\beta|) = 0.013^\circ$.

\begin{figure}[!h]
  \centering
  \includegraphics[width=0.55\textwidth]{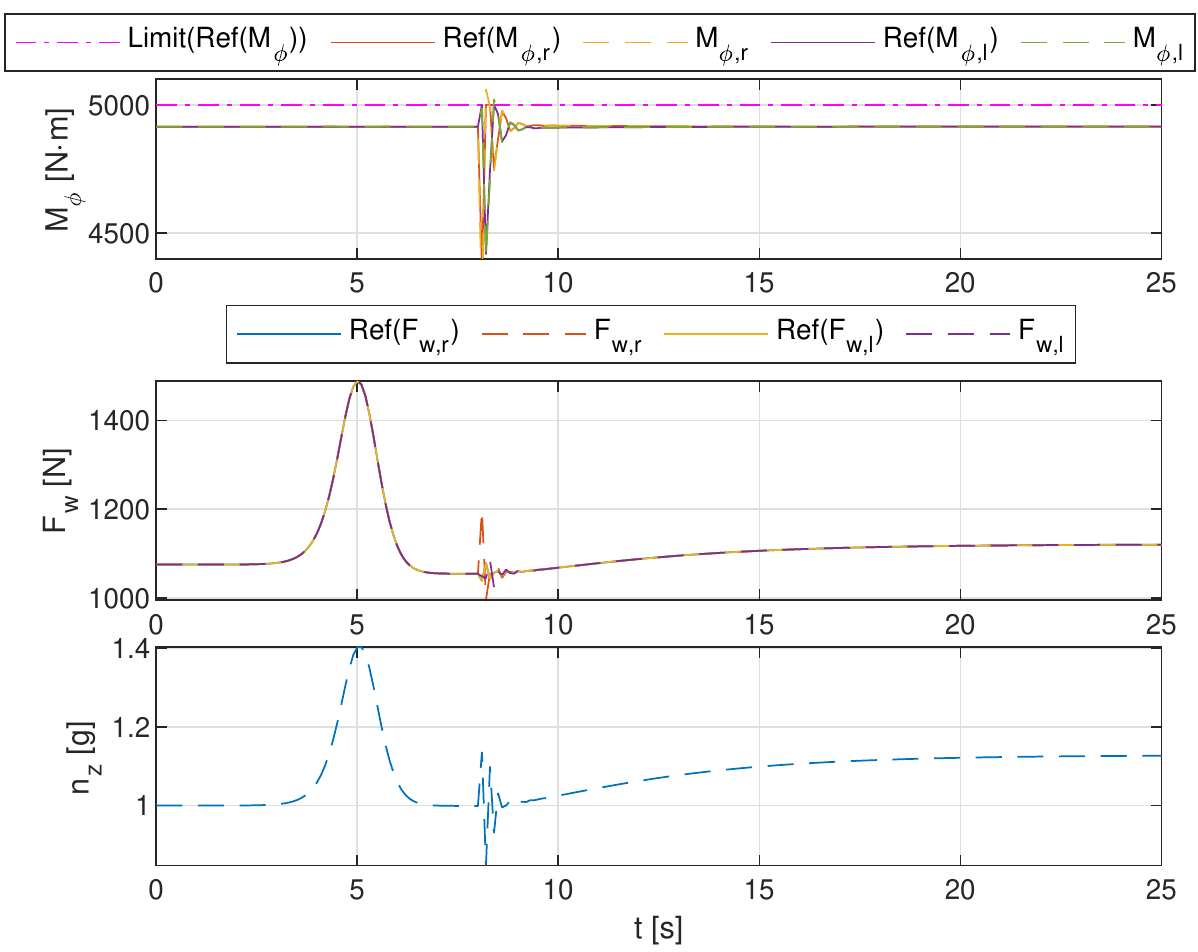}
  \caption{Responses of the wing-root bending moments, shear forces, and load factor in a spiral maneuver.}
  \label{track_sprial_loads}
\end{figure}

The load responses of the aircraft during the spiral maneuver are shown in Fig.~\ref{track_sprial_loads}. When $t\in[0,8]$~s, only symmetrical states are excited, thus $M_{\phi,r}$ and $M_{\phi,l}$ are maintained at their trimmed values, while $F_{w,r}, ~F_{w,l}$ are driven to realize the $\gamma$ tracking command (Eq.~\eqref{eq_gamma}). When $t\in[8,25]~s$, the aircraft starts entering the spiral. Similar to the responses in the sharp-roll maneuver illustrated in Fig.~\ref{sharp_roll_loads}, in case one of ${M}_{\phi,r},~{M}_{\phi,l}$ reaches $\overline {M}_{\phi}^{\text{ref}}$, the exceeded command is automatically allocated to the other wing.

\begin{figure}[!h]
\centering
\includegraphics[width=0.55\textwidth]{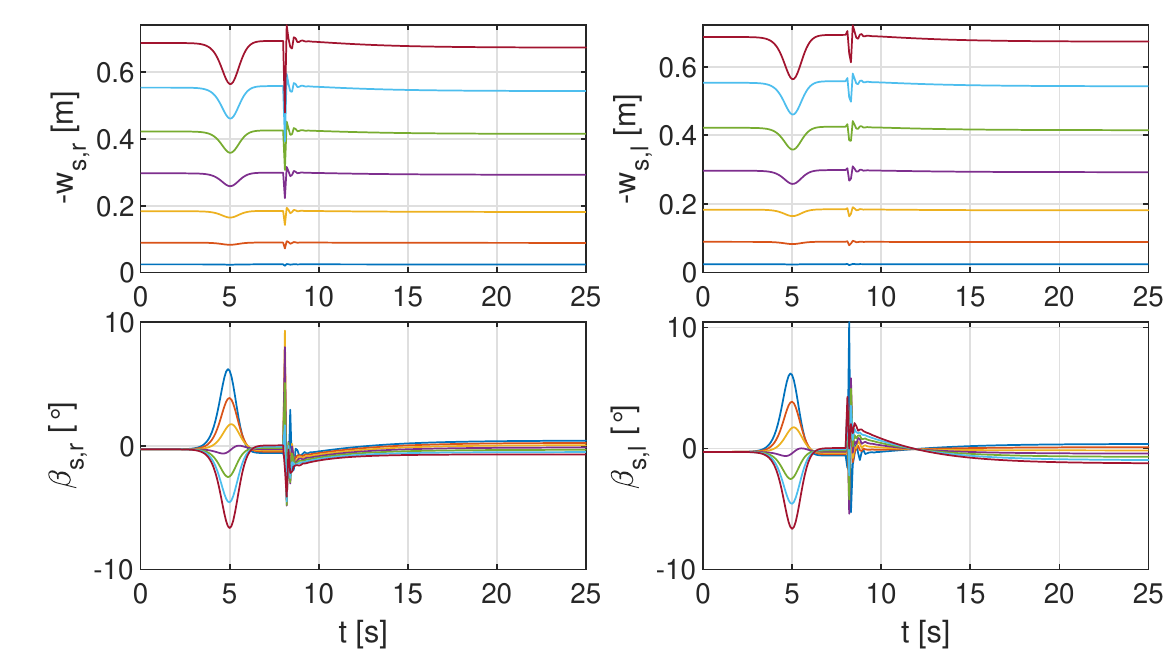}
  \caption{Responses of the wing bending displacements and flap deflections in a spiral maneuver.}
  \label{track_sprial_wing}
\end{figure}

In this spiral maneuver, the responses of the wing bending displacements and flap deflections are shown in Fig.~\ref{track_sprial_wing}. It can be seen when $t\in[0,8]~s$, the flap deflections are the same, while when $t\in[8,25]~s$, the flap deflections are re-allocated for roll tracking and MLA purposes. 
\subsubsection{Position Control With Maneuver Load Alleviation}
\label{subsection_3D}
\begin{figure}[!h]
  \centering
  \includegraphics[width=0.5\textwidth]{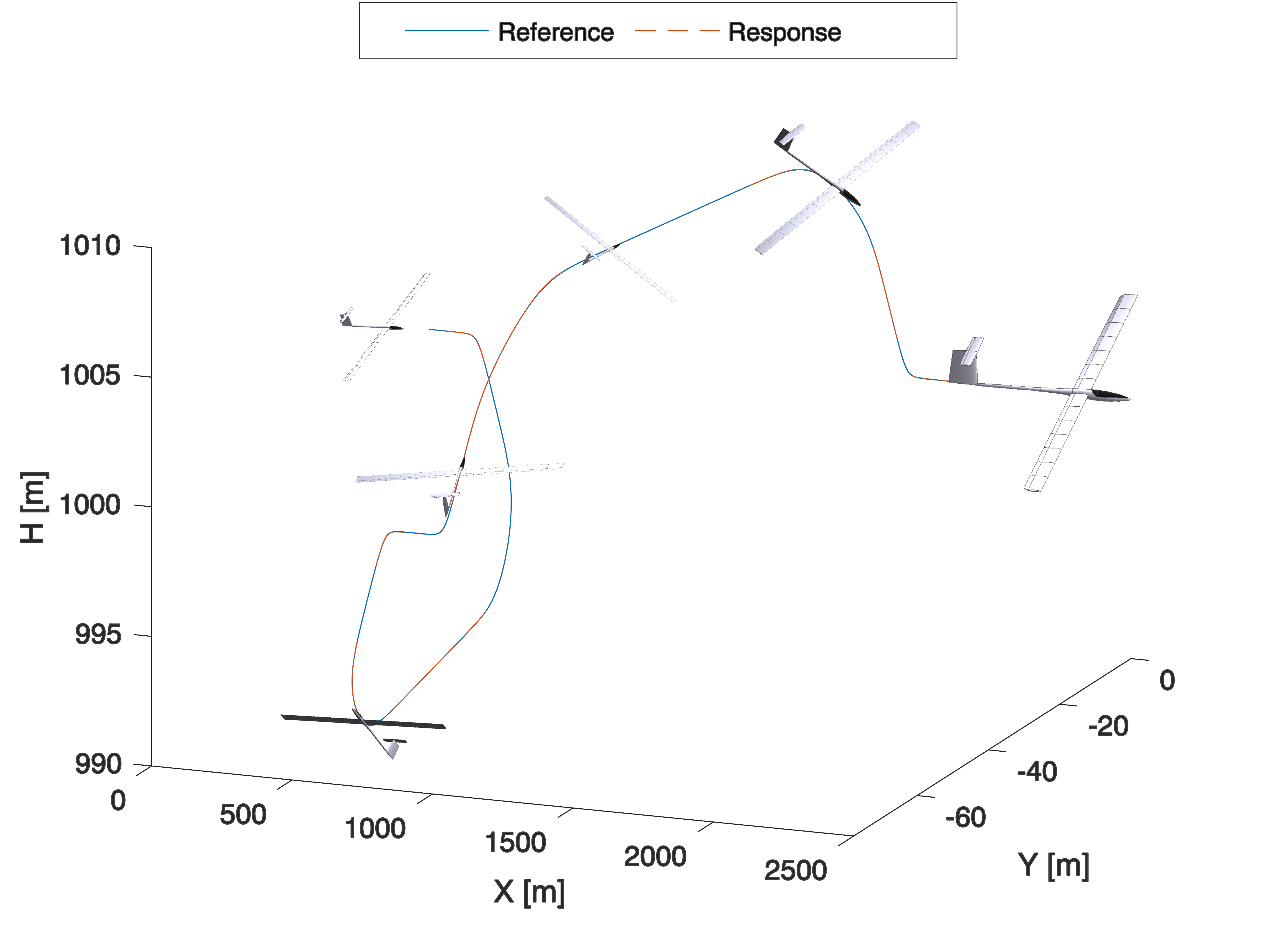}
  \caption{The flexible aircraft 3D trajectory tracking responses.}
  \label{track_3D_XYZ}
\end{figure}

\begin{figure}[!h]
  \centering
  \includegraphics[width=0.65\textwidth]{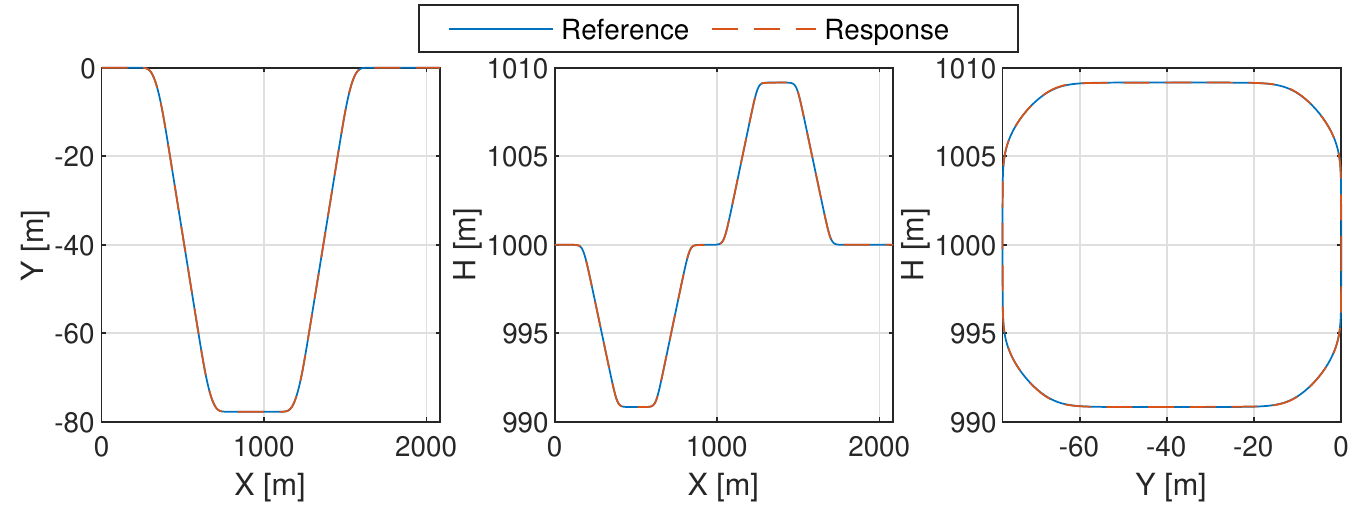}
  \caption{Three views of the 3D trajectory tracking performance.}
  \label{track_XYZ}
\end{figure}

A more complex 3D maneuver with strongly coupled longitudinal and lateral motions will be tested in this subsection. The trajectory references and responses are illustrated in Figs.~\ref{track_3D_XYZ} and~\ref{track_XYZ}. The proposed controller can minimize the trajectory errors, with $\text{max}(|X - X^{\text{ref}}|) < 1.4$~m, $\text{max}(|Y - Y^{\text{ref}}|) < 0.019$~m and $\text{max}(|H - H^{\text{ref}}|) < 0.023$~m during this maneuver.
 
\begin{figure}[!h]
  \centering
  \includegraphics[width=0.9\textwidth]{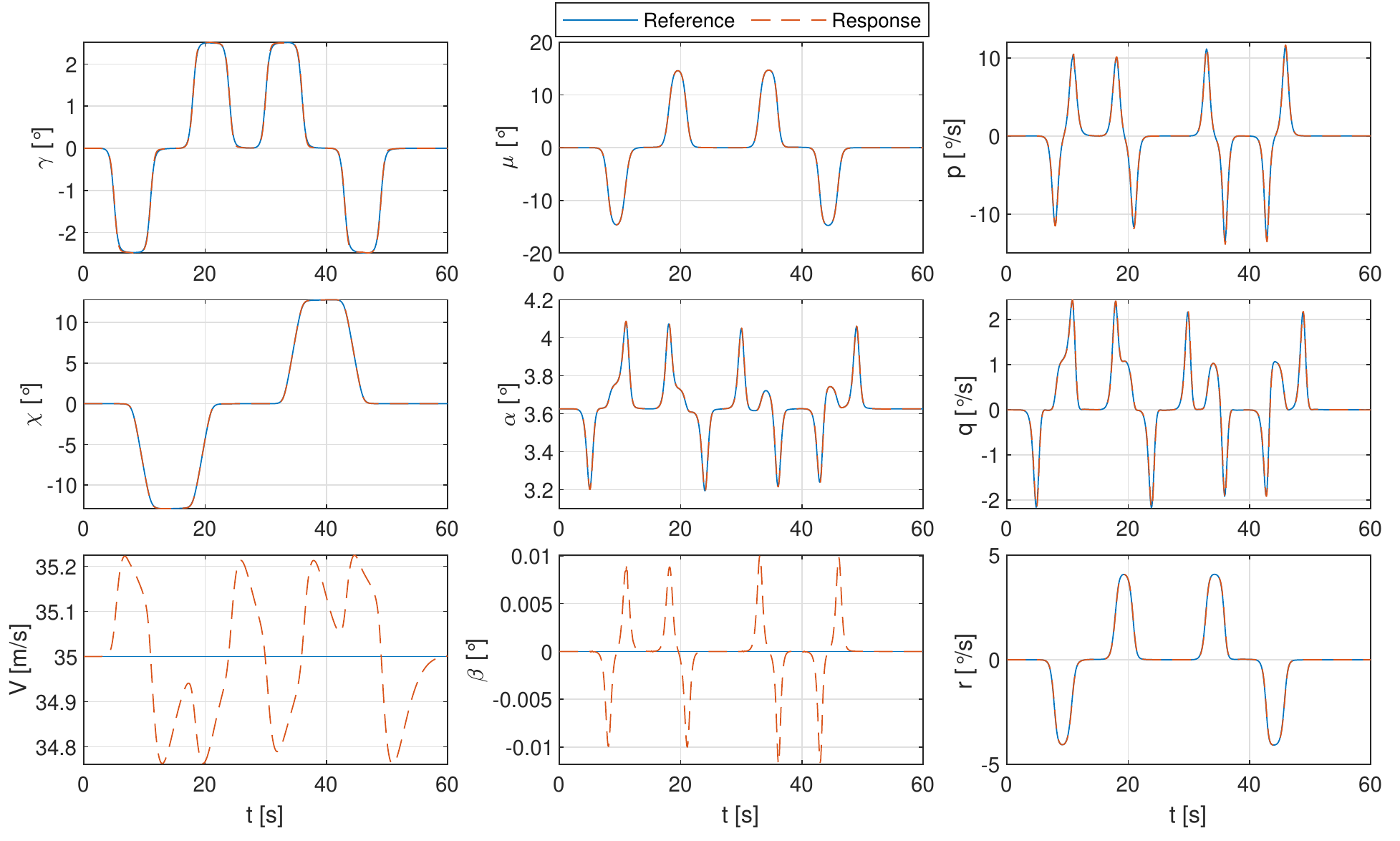}
  \caption{Responses of flight path angles, attitude angles, and angular rates in an aircraft 3D maneuver.}
  \label{track_miu_ab}
\end{figure}

The responses of rigid-body states are shown in Fig.~\ref{track_miu_ab}. Although the lateral and longitudinal control commands are coupled in this case, the proposed nonlinear controller can decouple the resulting dynamics through the control effectiveness inversion. As a result, the pitch, roll, and yaw channels are governed by their own designed tracking error dynamics, which are then realized by the virtual controls (Eq.~\eqref{eq_u_ibsmc}). In Fig.~\ref{track_miu_ab}, $\max(|e_\gamma|)= 0.0736^\circ$, and $\max(|e_\chi|)= 0.0604^\circ$.

\begin{figure}[!h]
  \centering
  \includegraphics[width=0.55\textwidth]{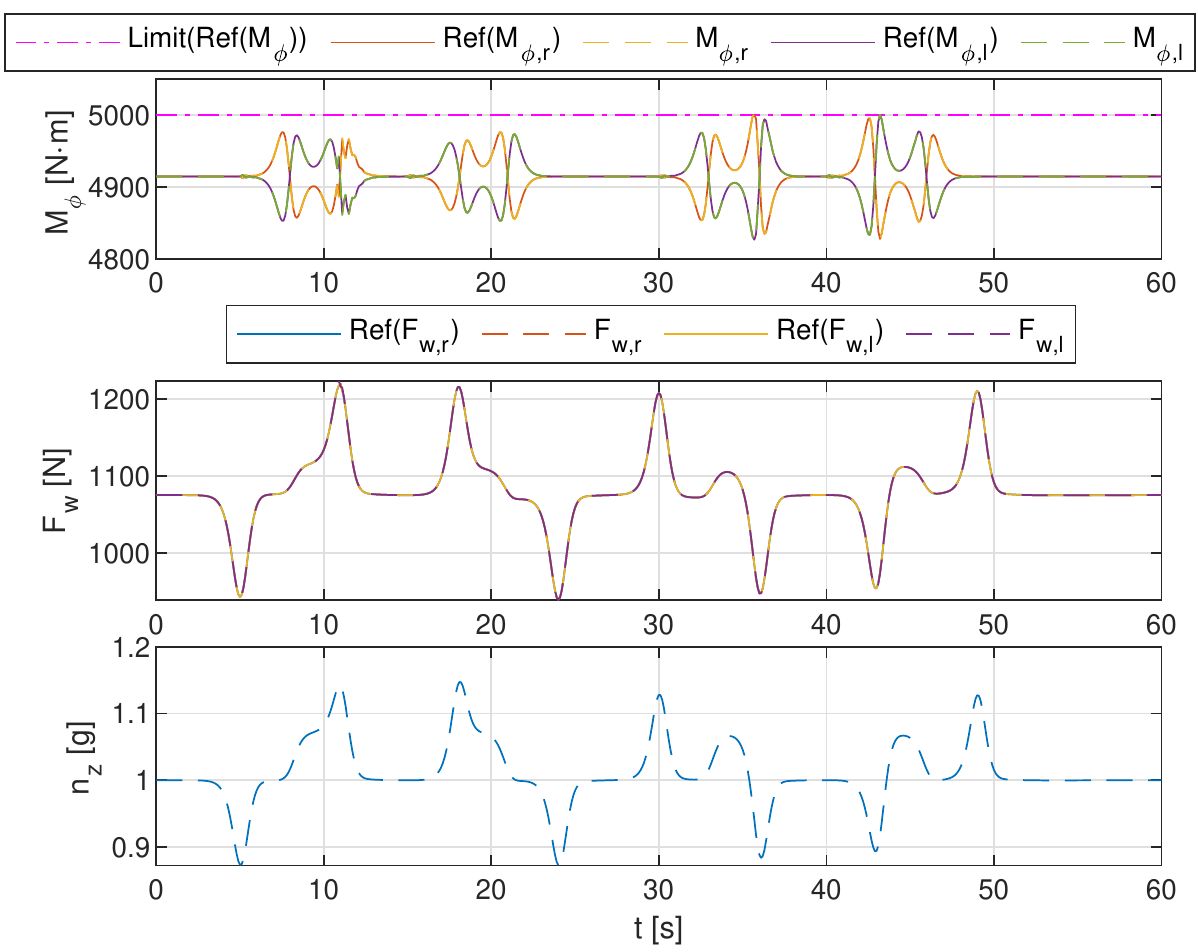}
  \caption{Responses of the wing-root bending moments, shear forces, and load factor in a 3D maneuver.}
  \label{track_loads}
\end{figure}

\begin{figure}[!h]
  \centering
  \includegraphics[width=0.55\textwidth]{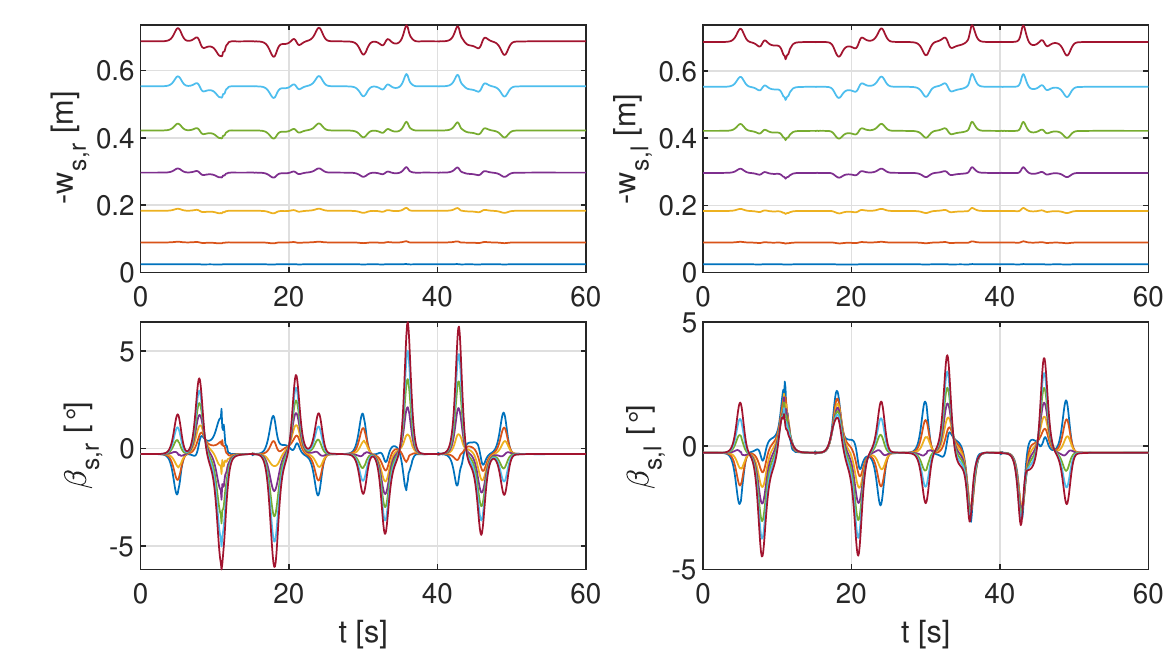}
  \caption{Responses of the wing bending displacements and flap deflections in a 3D maneuver.}
  \label{track_wing}
\end{figure}

The load responses are shown in Fig.~\ref{track_loads}. The differences between the right and left wing root bending moments are necessary for achieving lateral maneuvers (Sec.~\ref{sub_set_attitude}). By virtue of the MLA algorithm, the reference limit $\overline {M}_{\phi}^{\text{ref}} $ is not exceeded. The shear force commands are needed for achieving flight path angle command, and they have similar frequency component with the load factor $n_z$. Figure~\ref{track_wing} shows the wing displacements and flap deflections. During this maneuver, the maximum wing-tip displacement equals 0.74~m, which only deviates 0.06~m from its nominal value. The flap deflections are smooth, and are within the $\pm 30$~deg limit.
\subsection{Gust Load Alleviation}
\label{sub_GLA}
\begin{figure}[!h]
  \centering
  \includegraphics[width=0.55\textwidth]{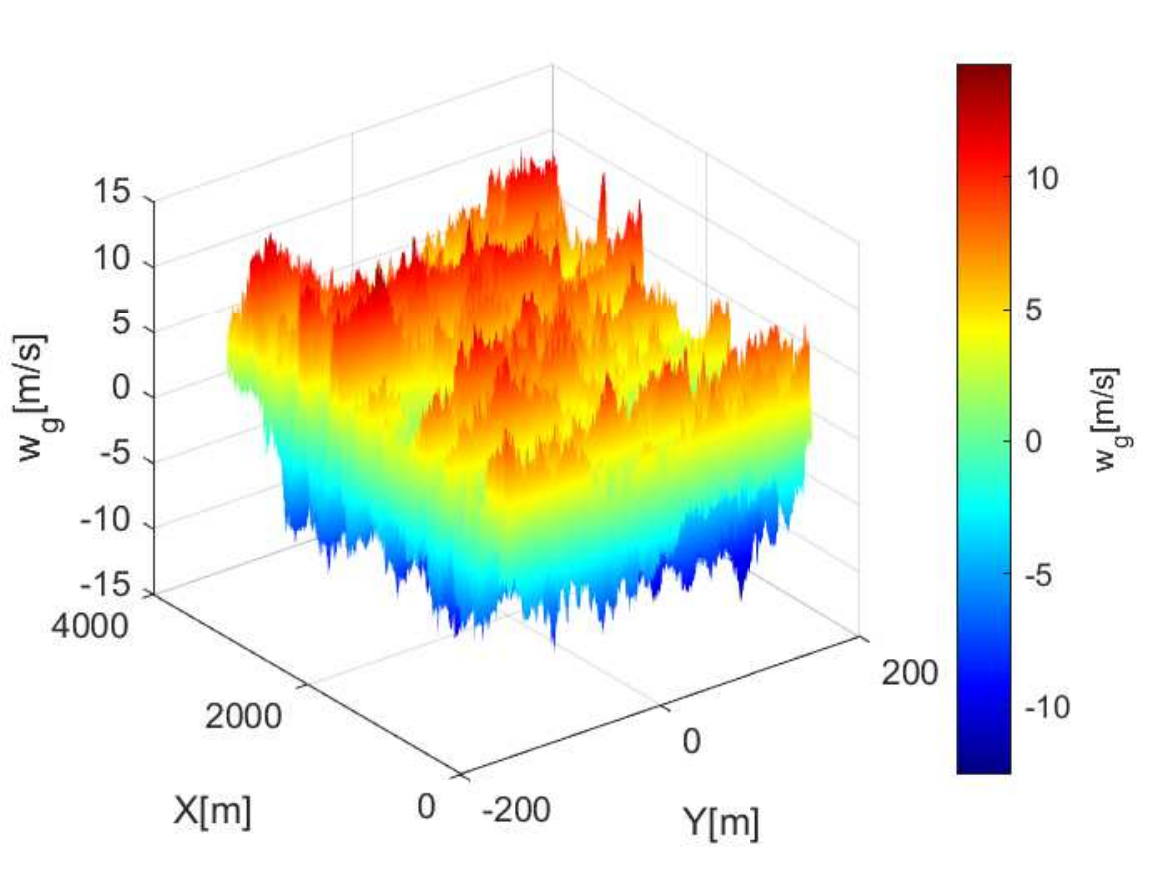}
  \caption{A 2D severe von K\'arm\'an turbulence field ($L_g = 762$~m, $\sigma =6$~m/s).}
  \label{severe_turb}
\end{figure}

In this section, the gust load alleviation performance of the designed controller will be evaluated. The aircraft is commanded to maintain at a cruise condition: $V= 35 \text{m/s}, ~H = 1000\text{m},~\gamma = \chi = 0$. The von K\'arm\'an turbulence model is adopted in this paper. Using two-dimensional Fourier transforms, the von K\'arm\'an spectrum can be realized in a two-dimensional spatial domain~\cite{Wang2019c}. Realizing a turbulence spectrum in spatial domain has several benefits: first of all, it allows the consideration of gust penetration effect~\cite{Etkin2014a}. Moreover, the span-wise gust velocity variations are modeled, which are important for high aspect-ratio aircraft. Furthermore, in a spatial turbulence field, the aircraft maneuvers are not constrained in the symmetric plane; multi-dimensional maneuvers are allowed. In addition, given the aircraft ground velocity, the spatial domain turbulence field can be easily embedded in time-domain simulations. Fig.~\ref{severe_turb} shows a severe von K\'arm\'an turbulence field with the turbulence scale length $L_g = 762$~m, and the turbulence intensity $\sigma =6$~m/s.

The control objective is to maintain the cruise flight, while alleviating the additional loads induced by the turbulence field. To demonstrate the effectiveness of the controller, it is compared to the open-loop condition. However, as discussed in subsection~\ref{sec_results}, the phugoid mode of this flexible aircraft is unstable. For fair comparisons, the ``open-loop'' responses shown in the following figures represent the aircraft with stabilized phugoid mode (via velocity feedback control). Even so, the elevator and rudder deflection angles are set to zero in the ``open-loop'' case, while the hinge actuation moments of the flaps are all set to zero. 

\begin{figure}[!h]
  \centering
  \includegraphics[width=0.55\textwidth]{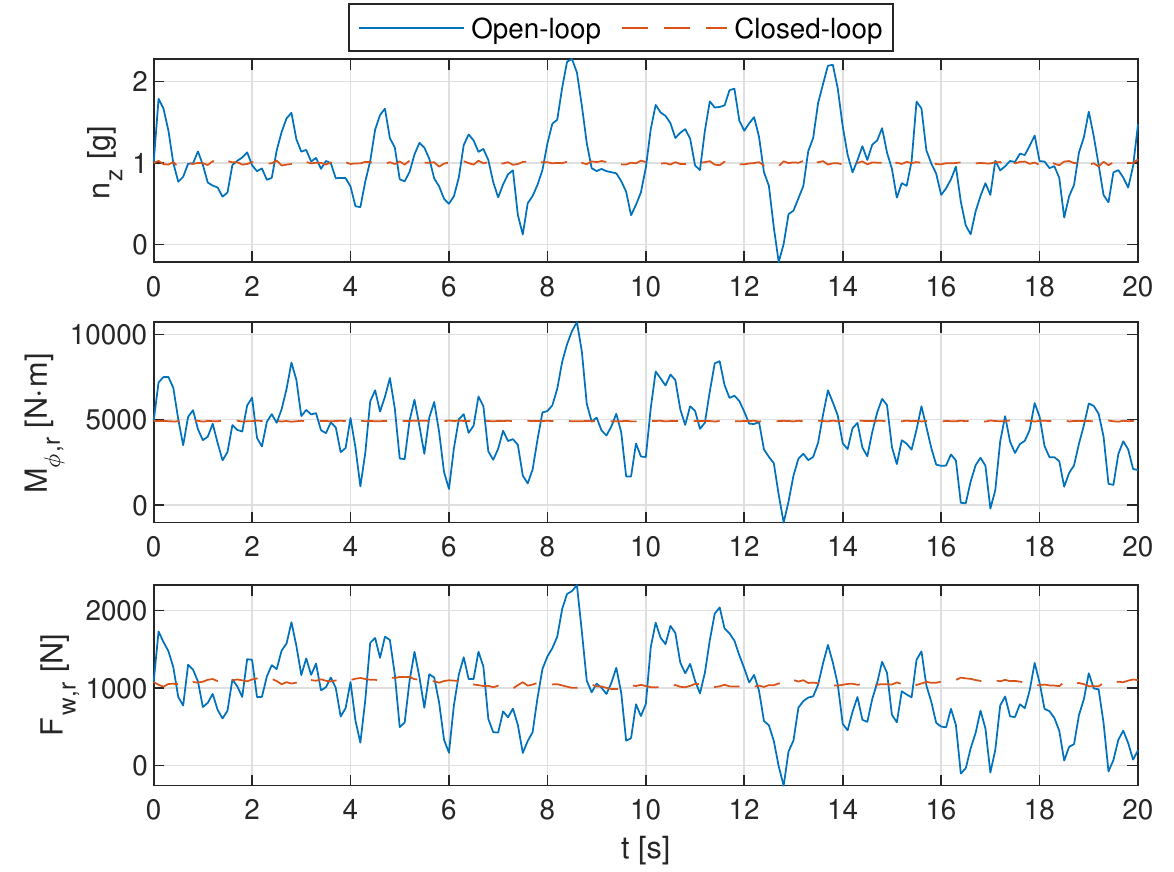}
  \caption{Responses of the load factor, root bending moment, and shear force in a von K\'arm\'an turbulence field.}
  \label{severe_turb_loads}
\end{figure}

\begin{figure}[!h]
  \centering
  \includegraphics[width=0.55\textwidth]{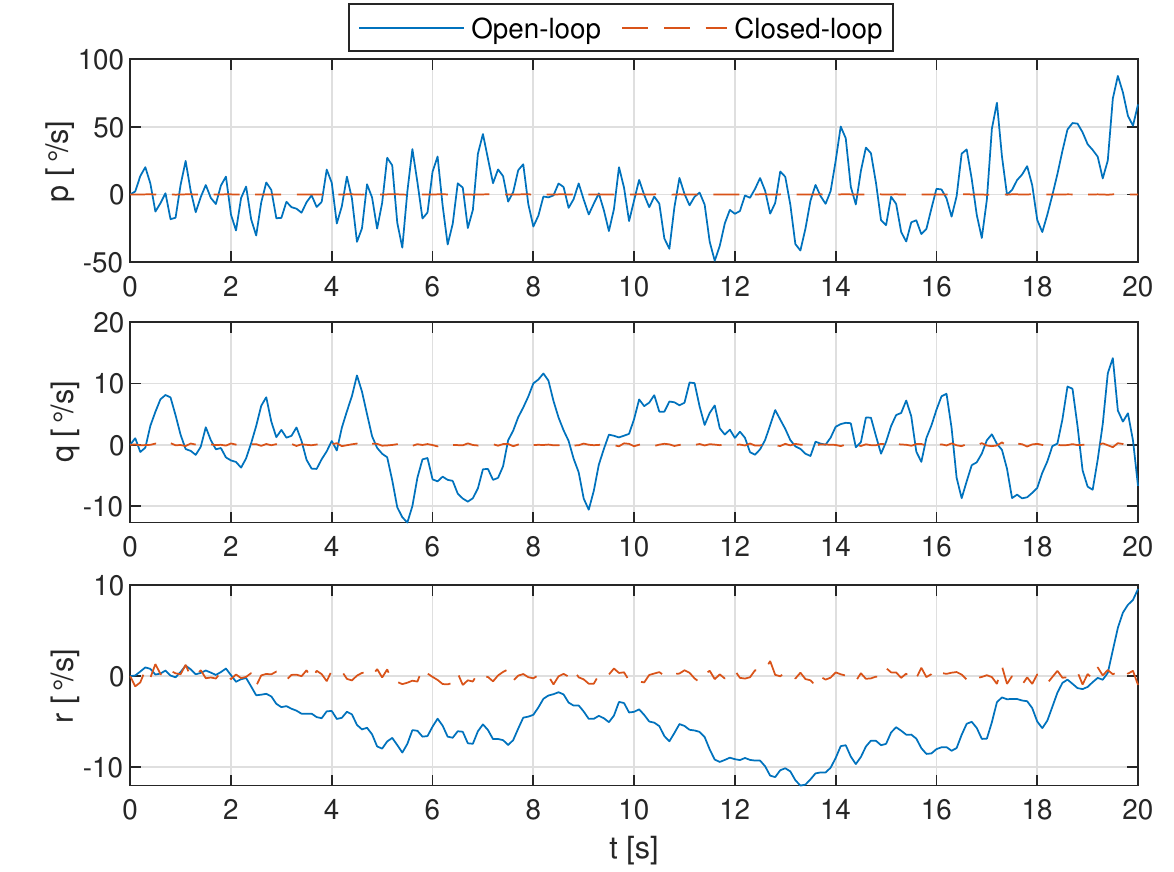}
  \caption{Angular velocity responses in a 2D von K\'arm\'an turbulence field.}
  \label{severe_turb_pqr}
\end{figure}

The responses of the right-wing root bending moment and shear force, and the load factor are shown in Fig.~\ref{severe_turb_loads}, from which it can be clearly seen the proposed controller effectively alleviates the load. To be specific, $\text{rms}(n_z- n_z^*)$ is reduced from 0.439~g to 0.0159~g (by 96.38\%), $\text{rms}(M_{\phi,r}- M_{\phi,r}^*)$ is alleviated from 2057~N$\cdot$m to 16.45~N$\cdot$m (by 99.20\%), and $\text{rms}(F_{w,r}- F_{w,r}^*)$ is diminished from 511.3~N to 38.68~N (by 92.43\%). The superscript $(\cdot)^*$ denotes the trimmed value. Moreover, the responses of angular rates are shown in Fig.~\ref{severe_turb_pqr}. In the ``open-loop'' case, the roll rate and pitch rate are non-zero because the local angles of attack induced by the 2D spatial turbulence field are different on each strip. For the high-aspect ratio aircraft model used in this paper, the turbulence field has strongest influence on the roll channel, with $\text{rms}(p) = 23.5^\circ$/s in the ``open-loop'' case, which is reduced to $0.0812^\circ$/s using the proposed controller (reduced by 99.65\%). Besides, this control architecture diminishes $\text{rms}(q) = $ from 5.31$^\circ$/s to 0.134$^\circ$/s (by 97.48\%), and reduces $\text{rms}(r) = $ from 5.99$^\circ$/s to 0.505$^\circ$/s (by 91.57\%). As discussed in~\cite{Wang2017}, the load factor is tightly linked with passenger ride comfort, thus by alleviating $n_z$, the ride comfort can also be improved.

\begin{figure}[!h]
  \centering
  \includegraphics[width=0.55\textwidth]{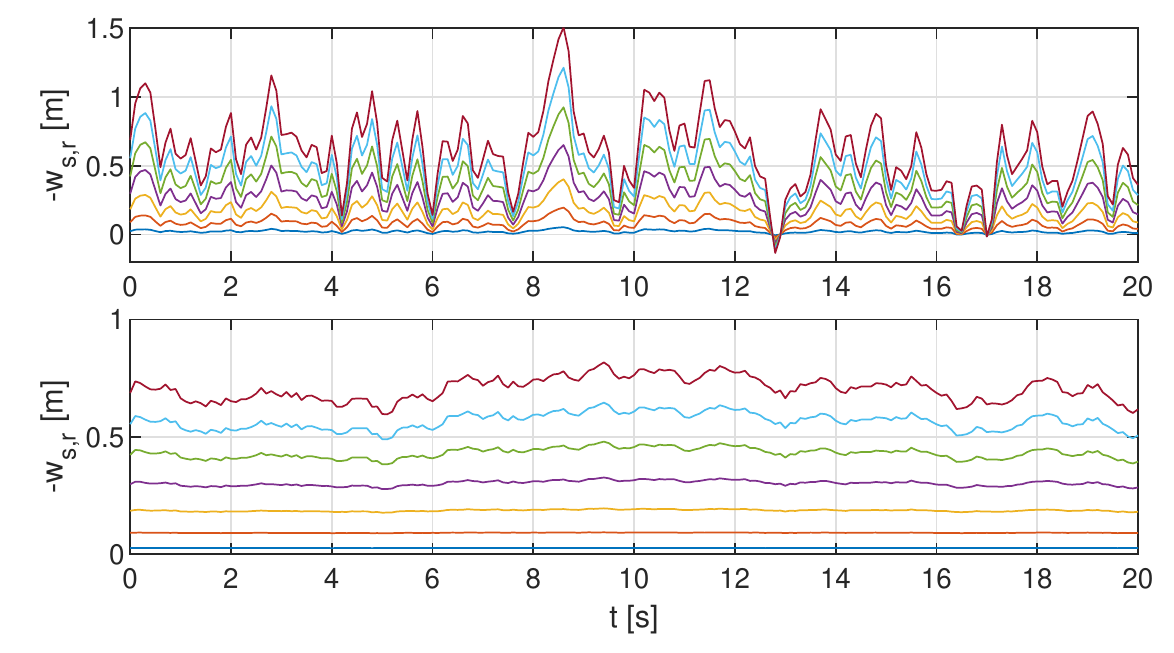}
  \caption{Displacements of the seven structural nodes on the right wing (top: open-loop, bottom: closed-loop).}
  \label{severe_turb_wing_dis}
\end{figure}

\begin{figure}[!h]
  \centering
  \includegraphics[width=0.55\textwidth]{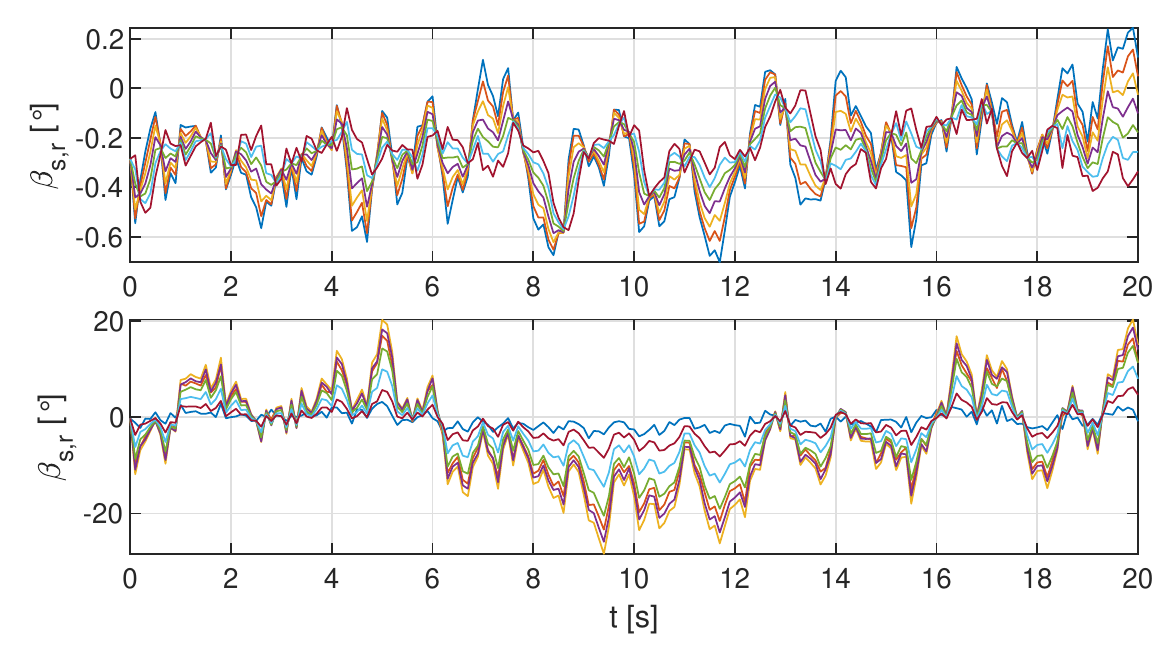}
  \caption{Deflections of the seven flaps on the right wing (top: open-loop, bottom: closed-loop).}
  \label{severe_turb_flap}
\end{figure}

The wing displacements in the ``open-loop'' and closed-loop cases are illustrated in Fig.~\ref{severe_turb_wing_dis}. It can be seen that the structural motions are damped out by the proposed controller. The maximum deviation of the wing tip is reduced from 1.5~m to 0.81~m (by 46.00\%). The flap deflections are shown in Fig.~\ref{severe_turb_flap}. As mentioned in the preceding texts, the ``open-loop'' case gives zero hinge moments, but the flaps can still move due to their inertia and stiffness couplings with the main wing structure. When the control loop is closed, the flaps deflect in oppose to the turbulence profile, and modify the lift distributions at the same time. The deflection angles are within the limit of $\pm 30$~deg in this severe turbulence field.


\section{Conclusions}
\label{sec_conclusions}
In this paper, a nonlinear control architecture for flexible aircraft trajectory tracking and load alleviation is proposed. To begin with, this paper derives the kinematics and dynamics of the free-flying flexible aircraft, which capture both the inertial and aerodynamic couplings between the rigid-body and structural degrees of freedom. The dynamic model is derived in a modular approach, which provides a convenient way to make an existing clamped-wing aeroservoelastic model free-flying.

Based on the flexible aircraft model, a four-loop cascaded control architecture is proposed. The position control loop is designed using nonlinear dynamic inversion, which provides references to the flight path control loop. Because model uncertainties and disturbances present, the flight path control loop adopts the incremental sliding mode control law, whose reduced model dependency and enhanced robustness are demonstrated theoretically. Moreover, the incremental backstepping sliding mode control is used in the attitude control loop. Based on Lyapunov methods, this loop is proven to be stable under disturbance perturbations. Furthermore, a novel load reference generator is proposed, which distinguishes the loads that are necessary to perform maneuvers from the excessive loads. The load references are then tracked by the inner-loop multi-objective wing control. At the same time, the excessive loads (no matter induced by maneuvers or gusts) are naturalized by control surface deflections. Since the proposed control architecture exploits the control redundancy, the load alleviation is achieved without affecting the aircraft command tracking performance. 

The effectiveness of the proposed control architecture is validated by numerical simulations. It is demonstrated that the loads during sudden pull-up and sharp roll maneuvers can be alleviated without influencing rigid-body command tracking performance. Moreover, a simulation case that the flexible aircraft flies through a severe 2D spatial von K\'arm\'an turbulence field verifies the robustness of the controller to model uncertainties and external disturbances. Finally, the flexible aircraft is commanded to follow a 3D trajectory in a spatial turbulence field. Simulation results show that both the maneuver and gust loads are effectively alleviated without affecting the trajectory tracking performance. The extension of the method to highly flexible aircraft with nonlinear beam structures will be explored in future work. 

\section*{Acknowledgement}
The first author would like to thank Lan Yang, Paul Lancelot, and Yang Meng for their help on the dynamic model. 
\bibliography{glider_conf}

\end{document}